\documentstyle[12pt,a4,cite,psfig]{article}

\makeatletter
 
\renewenvironment{thebibliography}[1]
{\small
 \begin{list}{[\arabic{enumi}]}
 {\usecounter{enumi} \setlength{\parsep}{6pt}
  \setlength{\itemsep}{3pt} \settowidth{\labelwidth}{[#1]}
  \settowidth{\leftmargin}{[100]}
  \sloppy}}
 {\end{list}}
  
\renewcommand{\theequation}{\arabic{equation}}
 
\newtoks\@stequation

\def\subequations{\refstepcounter{equation}%
  \edef\@savedequation{\the\c@equation}%
  \@stequation=\expandafter{\theequation}
  \edef\@savedtheequation{\the\@stequation}
  \edef\oldtheequation{\theequation}%
  \setcounter{equation}{0}%
  \def\theequation{\oldtheequation\alph{equation}}}

\def\endsubequations{%
  \ifnum\c@equation < 2 \@warning{Only \the\c@equation\space subequation
    used in equation \@savedequation}\fi
  \setcounter{equation}{\@savedequation}%
  \@stequation=\expandafter{\@savedtheequation}%
  \edef\theequation{\the\@stequation}%
  \global\@ignoretrue}

\def\eqnarray{\stepcounter{equation}\let\@currentlabel\theequation
\global\@eqnswtrue\m@th
\global\@eqcnt\z@\tabskip\@centering\let\\\@eqncr
$$\halign to\displaywidth\bgroup\@eqnsel\hskip\@centering
     $\displaystyle\tabskip\z@{##}$&\global\@eqcnt\@ne
      \hfil$\;{##}\;$\hfil
     &\global\@eqcnt\tw@ $\displaystyle\tabskip\z@{##}$\hfil
   \tabskip\@centering&\llap{##}\tabskip\z@\cr}

\def\dlinepattern#1#2{%
\ifdim#2<#1
   \errmessage{the 1st argument is less than the 2nd argument.}%
\else
   \gdef\dline@solid{#1}\gdef\dline@period{#2}%
\fi}

\def\dline#1{\@dline[#1]}
\def\@dline[#1-#2]{\noalign{\global\@dla#1\relax
\global\advance\@dla\m@ne
\ifnum\@dla>\z@\global\let\@gtempa\@dlinea\else
  \global\let\@gtempa\@dlineb\fi
\global\@dlb#2\relax
\global\advance\@dlb-\@dla}\@gtempa
\noalign{\vskip-\arrayrulewidth}}

\def\@dlinea{\multispan\@dla&\multispan\@dlb
\unskip\cleaders\hbox to \dline@period
{\hss\rule{\dline@solid}{\arrayrulewidth}\hss}\hfill\cr}
 
\newcount\@dla
\newcount\@dlb

\def\@dlineb{\multispan\@dlb
\unskip\cleaders\hbox to \dline@period
{\hss\rule{\dline@solid}{\arrayrulewidth}\hss}\hfill\cr}

\dlinepattern{2pt}{5pt}

\long\def\@makecaption#1#2{%
   \vskip 10\p@
   \setbox\@tempboxa\hbox{{\small\bf #1}\ \ {\small #2}}%
   \ifdim \wd\@tempboxa >\hsize
       {\small\bf #1}\ \ {\small #2}\par
     \else
       \hbox to\hsize{\hfil\box\@tempboxa\hfil}%
   \fi}

\makeatother

\newcommand\commentout[1]{}
\newcommand\pr{\hspace{\parindent}}

\newcommand\textfrac[2]{{\textstyle\frac{#1}{#2}}}

\def\simgt{\rlap{\lower 3.5 pt \hbox{$\mathchar \sim$}}%
           \raise 1pt \hbox {$>$}}
\def\simlt{\rlap{\lower 3.5 pt \hbox{$\mathchar \sim$}}%
           \raise 1pt \hbox {$<$}}

\def\gev{{\,\rm GeV}}

\def\tev{{\,\rm TeV}}

\def\msbar{\overline{\rm MS }}	

\def\O{{\cal O}}

\def\mv{m_V^{}}
\def\mz{m_Z^{}}
\def\mw{m_W^{}}
\def\mh{m_H^{}}
\def\mmv{m_V^2}
\def\mmz{m_Z^2}
\def\mmw{m_W^2}

\def\ebar{\bar{e}}
\def\sbar{\bar{s}}

\def\gzbar{\bar{g}_Z}
\def\gwbar{\bar{g}_W}
\def\ehat{\hat{e}}
\def\shat{\hat{s}}
\def\chat{\hat{c}}
\def\gzhat{\hat{g}_Z}
\def\ghat{\hat{g}}
\def\msbar{\overline{\rm MS}}

\def\pibar{\overline{\Pi}}

\def\delb{\bar{\delta}_{b}}
\def\delg{\bar{\delta}_{G}^{}}
\def\zbb{Zb_L^{}b_L^{}}

\def\tableofdata{%
 \def\afb{A_{\rm FB}}                
 \begin{center}
 \begin{tabular}{|r|c|cccc|} 
 \hline
 \multicolumn{1}{|l|}{measurement}
 & data  &             
 \multicolumn{4}{|c|}{SM prediction 
                      ($\alpha_s\!=0.116,\:\delta_\alpha\!=0$)}\\
 \hline
         $m_t$ (GeV)   &               
 &    150 &    150 &    175 &    175\\
         $m_H^{}$ (GeV)&               
 &    100 &   1000 &    100 &   1000\\
 \hline
\multicolumn{1}{|l|}{{\bf LEP}\cite{lep94_glasgow}} & & & & & \\
\multicolumn{1}{|l|}{line shape:} & & & & & \\
$\mz$(GeV) & 91.1888$\pm$0.0044 &
\multicolumn{4}{|c|}{(input)} \\ 
$\Gamma_Z$(GeV) &  2.4974 $\pm$ 0.0038 
 & 2.4906 & 2.4823 & 2.4965 & 2.4877\\
$\sigma_h^0$(nb)&   41.49 $\pm$   0.12 
 &  41.47 &  41.48 &  41.48 &  41.49\\
$R_\ell\equiv\Gamma_h/\Gamma_\ell$ &  20.795 $\pm$  0.040 
 & 20.738 & 20.714 & 20.729 & 20.706\\
$\afb^{0,\ell}$ &  0.0170 $\pm$ 0.0016 
 & 0.0153 & 0.0132 & 0.0167 & 0.0145\\[2mm]
\multicolumn{1}{|l|}{$\tau$ polarization:} & & & & & \\
$A_\tau$        &   0.143 $\pm$  0.010 
 &  0.142 &  0.132 &  0.148 &  0.138\\
$A_e$           &   0.135 $\pm$  0.011 
 &  0.142 &  0.132 &  0.148 &  0.138\\[2mm]
\multicolumn{1}{|l|}{$b$ and $c$ quark results:} & & & & & \\
$R_b\equiv\Gamma_b/\Gamma_h$&  0.2202 $\pm$ 0.0020 
 & 0.2165 & 0.2166 & 0.2157 & 0.2157\\
$R_c\equiv\Gamma_c/\Gamma_h$&  0.1583 $\pm$ 0.0098 
 & 0.1718 & 0.1717 & 0.1721 & 0.1720\\
$\afb^{0,b}$    &  0.0967 $\pm$ 0.0038 
 & 0.0994 & 0.0923 & 0.1038 & 0.0965\\
$\afb^{0,c}$    &  0.0760 $\pm$ 0.0091 
 & 0.0710 & 0.0655 & 0.0744 & 0.0688\\[2mm]
\multicolumn{1}{|l|}{{\bf SLC}\cite{alr94}} & & & & & \\
$A_{\rm LR}^0$  &  0.1637 $\pm$ 0.0075 
 & 0.1420 & 0.1320 & 0.1482 & 0.1380\\
\hline
    $\chi^2/({\rm d.o.f.})$        &         
 & 19.8/11 &  50.7/11 &  17.6/11 &  31.8/11\\
 \hline
 \end{tabular}
 \end{center}
}

\def\fitofgzbsbnnu{%
  \begin{subequations}
  \begin{eqnarray}
    &&
    \left.
    \begin{array}{lll}
     \gzbar^2(\mmz) &\!\!=\!\!& 0.55663
       -0.00034\,\textfrac{\alpha_s +1.6\,\delb -0.100}{0.005}
       \pm 0.00113
   \\[1mm]
     \sbar^2(\mmz)  &\!\!=\!\!& 0.23052
       +0.00007\,\textfrac{\alpha_s +1.6\,\delb -0.100}{0.005}
        \pm 0.00042
   \\[1mm]
     N_\nu &\!\!=\!\!& 3.003
       -0.0068\,  \textfrac{\alpha_s +1.6\,\delb -0.100}{0.005}
        \pm 0.021 
    \end{array}
    \right\}
    \;
     \rho_{\rm corr} = 
     \left(\begin{array}{rrr}
            1 &   0.19 &  -0.63 \\[1mm]
              &   1    &   0.03 \\[1mm]
              &        &   1
           \end{array}\right) \,,
    \nonumber \\ &&
    \\
    & & \qquad
      \chi^2_{\rm min} = 11.2 
        +\biggl(\frac{\alpha_s +1.60\,\delb -0.1099}{0.0060} \biggr)^2
        +\biggl(\frac{\delb +0.0015}{0.0046}\biggr)^2 .
  \end{eqnarray}
  \end{subequations}
}

\def\fitofnnu{%
  \begin{eqnarray}
      N_\nu &=&  2.993
         -0.0128\,\textfrac{\gzbar^2(\mmz) -  0.55550}{0.00101}
         +0.0035\,\textfrac{\sbar^2(\mmz)  -  0.23068}{0.00042}
      \nonumber \\
      &&
         -0.0098\,\textfrac{\delb(\mmz)    +  0.0034 }{0.0026 }
         -0.0118\,\textfrac{\alpha_s       -  0.116  }{0.005  }
     \pm  0.016 \,,
  \end{eqnarray}
}

\def\fitofgzbsb{%
  \begin{subequations}
   \label{fitofgzbsb}
  \begin{eqnarray}
     & &
     \left.
     \begin{array}{ll}
     \gzbar^2(\mmz) &= 0.55673 
      -0.00056\,\textfrac{\alpha_s +1.6\,\delb -0.100}{0.005} \pm 0.00087
      \\
     \sbar^2(\mmz)  &= 0.23051 
      +0.00008\,\textfrac{\alpha_s +1.6\,\delb -0.100}{0.005} \pm 0.00042
     \end{array}
    \right\}\quad
   \rho_{\rm corr} = 0.28,
   \\ 
   & & \quad
   \chi^2_{\rm min} =  11.4
         +\biggl(\frac{\alpha_s +1.60\,\delb -0.1089}{0.0056}\biggr)^2
         +\biggl(\frac{\delb+0.0015}{0.0046}\biggr)^2\,,
   \label{fitofgzbsbchisq}
  \end{eqnarray}
  \end{subequations}
}

\def\tableofsbar{%
  \begin{center}
  \begin{tabular}{|l|c|r|c|}
    \hline
    measurements & extracted $\sbar^2(\mmz)$ & $\chi\;\;$ & 
    $\chi^2_{\rm min}/{\rm (d.o.f.)}$
    \\ \hline
      $A_{\rm FB}^{0,\ell} $&$ 0.2302 \pm 0.0009$
     & $-0.5$ & ---    \\
      $A_{\tau}            $&$ 0.2310 \pm 0.0013$
     & $ 0.3$ & ---    \\
      $A_e                 $&$ 0.2321 \pm 0.0014$
     & $ 1.0$ & ---    \\
      $A_{\rm LR}^0        $&$ 0.2284 \pm 0.0010$
     & $-2.4$ & ---    \\ 
      $A_{\rm FB}^{0,b}    $&$ 0.2316 \pm 0.0007$
     & $ 1.5$ & ---    \\
      $A_{\rm FB}^{0,c}    $&$ 0.2300 \pm 0.0021$
     & $ -0.3$ & --- 
    \\ \hline
      $ A_{\rm FB}^{0,\ell}, A_e, A_{\tau}, A_{\rm LR}^0
                           $&$ 0.2301 \pm 0.0005$
     & $ -1.1$ & $5.7/3$
    \\ \hline
      $A_{\rm FB}^{0,\ell}, A_{\tau}, A_e, A_{\rm LR}^0, 
       A_{\rm FB}^{0,b}, A_{\rm FB}^{0,c}
                           $&$ 0.2306 \pm 0.0004$
    & $0$ & $9.4/5$
    \\ \hline
  \end{tabular}
  \end{center}
}%
\def\bydgzb{\,\textfrac{\gzbar^2(\mmz)-0.55550}{0.00101}}
\def\bydsb{\,\textfrac{\sbar^2(\mmz)-0.23068}{0.00042}}
\def\bydalps{\,\textfrac{\alpha_s-0.116}{0.005}}
\def\byddelb{\,\textfrac{\delb + 0.0034}{0.0026}}
\def\tableofdelb{%
 \begin{center}
 \begin{tabular}{|c|c|c|c|c|}
    \hline
    measurements & $\langle\delb\rangle \pm \Delta\delb$
                 & $C(\gzbar^2)$ & $C(\sbar^2)$ & $C(\alpha_s)$ 
    \\ \hline
    $\Gamma_Z$  &$ -0.0036\pm 0.0041$&$ -0.00499$&$ +0.00118$&$ -0.00311$
    \\ 
    $\sigma_h^0$&$ -0.0106\pm 0.0142$&  ---      &$ +0.00021$&$ -0.00312$
    \\
    $R_\ell$    &$ -0.0036\pm 0.0037$&  ---      &$ +0.00065$&$ -0.00311$
    \\
    $R_b$       &$ +0.0011\pm 0.0051$&  ---      &$ -0.00004$&$ -0.00005$
    \\ \hline
    All         &$ -0.0034\pm 0.0023$&$ -0.00145$&$ +0.00065$&$ -0.00230$
    \\ \hline
 \end{tabular}
 \end{center}
}

\def\fitofalps{%
  \begin{eqnarray}
    \label{fitofalps}
     \alpha_s &=&  0.1150
          -0.0032\bydgzb +0.0015\bydsb -0.0042\byddelb\,\pm 0.0044\,.
        \quad\;
  \end{eqnarray}
}

\def\tableofalps{%
\begin{center}
\begin{tabular}{ccccccc}
\hline \vphantom{$\Big/$}
$m_t\,$(GeV) & $\mh\,$(GeV) & $\gzbar^2(\mmz)$ & $\sbar^2(\mmz)$ 
& $\delb(\mmz)$ & extracted $\alpha_s$ 
& $\chi^2_{\rm min}/{({\rm d.o.f})}$
\\
\hline
150 & 100 &0.55519 &0.23117 &$-0.0079$& $0.1253 \pm 0.0044$ &15.2/10 \\
150 &1000 &0.55408 &0.23243 &$-0.0079$& $0.1335 \pm 0.0044$ &34.7/10 \\
175 & 100 &0.55644 &0.23038 &$-0.0100$& $0.1218 \pm 0.0044$ &15.8/10 \\
175 &1000 &0.55527 &0.23167 &$-0.0100$& $0.1303 \pm 0.0044$ &21.1/10 \\
\hline
\end{tabular}
\end{center}
}

\def\tableoflencatmz{%
  \begin{center}
  \begin{tabular}{|c|ccrc|}
     \hline
     & $\gzbar^2(\mmz)$ & $\sbar^2(\mmz)$ & $\rho_{\rm corr}$
     & $\chi^2_{\rm min}/({\rm d.o.f.})$
     \\ \hline
     $\nu_\mu$--$f$ ($\nu_\mu$--$e$ $+$ $\nu_\mu$--$q$) &
          $0.5568\pm 0.0048$ & $0.2331\pm 0.0072$ & $0.75$  & 0.19/4
     \\
     $e$--$q$       (APV $+$ $e$--D) &
          $0.5583\pm 0.0170$ & $0.2188\pm 0.0093$ & $-0.62$ & 0.46/1
     \\ \hline
     All ($\nu_\mu$--$f$ + $e$--$q$) &
          $0.5533\pm 0.0037$ & $0.2266\pm 0.0047$ & $0.53$ & 2.22/7
     \\ \hline
  \end{tabular}
  \end{center}
}%
\def\fitofmw{%
  \begin{equation}
     \label{fitofmw}
	\gwbar^2(0) = 0.4225 -0.0031\,\frac{\delg-0.0055}{\alpha}
       \pm 0.0017\,.
  \end{equation}
}%

\def\chisqsm{
  \begin{subequations}
  \label{total_chisqsm}
  \begin{eqnarray}
     \chi^2_{\rm SM}(m_t,\mh,\alpha_s,\delta_\alpha)
     &=& \biggl(\frac{m_t -\langle m_t\rangle}{\Delta m_t} \biggr)^2
         +\chi^2_{H}(\mh,\alpha_s,\delta_\alpha)\,,
     \label{chisqsm}
  \end{eqnarray}
}
\def\fitofmt{%
  \begin{eqnarray}
     \label{fitofmt}
    \langle m_t \rangle &=& 164.9 +12.5\, \ln\frac{\mh}{100}
                            +1.0 \, \ln^2\frac{\mh}{100}
           -2.6\,\biggl(\frac{\alpha_s-0.116}{0.005}\biggr)
           -4.8\,\biggl(\frac{\delta_\alpha}{0.10}\biggr)\,,
     \label{fitofmtbest}
     \\
     \Delta m_t   &=& 9.0 - 0.07\, \ln \frac{\mh}{100}
          - \Bigl(0.24 -0.036\,\ln \frac{\mh}{100}\Bigr)\,
            \frac{m_t -175}{10}\,,
     \label{fitofmterror}
  \end{eqnarray}
}%
\def\chisqhsm{%
  \begin{eqnarray}
     \chi^2_H(\mh,\alpha_s,\delta_\alpha) &=&
      16.4 + \biggl(\frac{\delta_\alpha -0.30}{0.30} \biggr)^2
          + \biggl(\frac{\alpha_s-0.1240+0.0018\,\delta_\alpha}{0.0046}
            \biggr)^2
     \nonumber \\
     && \!\!\!\!\!
          - \biggl(\frac{\alpha_s-0.1376+0.046\,\delta_\alpha}{0.0133} 
                   \biggr) \ln \frac{\mh}{100}
          - \biggl(\frac{\alpha_s-0.1347}{0.028}\biggr)
                   \ln^2 \frac{\mh}{100}
          +\biggl(\frac{\delta_\alpha}{0.10}\biggr)^2.
     \nonumber \\ &&   \label{chisqminsm}
  \end{eqnarray}
  \end{subequations}
}%

\def\fitofstu{%
  \begin{subequations}
     \label{fitofstu}
  \begin{eqnarray}
  &&\!\! \left.
    \begin{array}{crrrr}
       S =&  \!\! -0.18 &
             \!\! -0.06\,\frac{\alpha_s+1.6\delb-0.100}{0.005} &
             \!\! +0.07\,\frac{\delta_\alpha}{0.10} &
             \!\! \pm 0.20
         \\[1mm]
       T =&  \!\!  0.93 &
             \!\! -0.13\,\frac{\alpha_s+1.6\delb-0.100}{0.005} & &
             \!\! \pm 0.20
         \\[1mm]
       U =&  \!\! -0.12 &
             \!\! +0.10\,\frac{\alpha_s+1.6\delb-0.100}{0.005} &
             \!\! +0.02\,\frac{\delta_\alpha}{0.10} &
             \!\! \pm 0.50 \\[1mm]
    \end{array}
    \right\} \quad
   \rho_{\rm corr} = \left(
      \begin{array}{rrr}
           1    &  0.84 & -0.08  \\[1mm]
                &  1    & -0.22  \\[1mm]
                &       &  1     \\[1mm]
      \end{array}
      \right),
   \nonumber \\ && \hspace{-5mm} \label{fitofstu_mean}
   \\
   & & \chi^2_{\rm min} =  14.4
         +\biggl(\frac{\alpha_s +1.60\,\delb -0.1091}{0.0055}\biggr)^2
         +\biggl(\frac{\delb+0.0015}{0.0046}\biggr)^2\,.
     \label{chisqofstu}
  \end{eqnarray}
  \end{subequations}
}%
\def\tableofotherdata{%
\begin{center}
 \begin{tabular}{|r|c|cccc|} 
 \hline
 \multicolumn{1}{|l|}{measurement}
 & data  &             
 \multicolumn{4}{|c|}{SM prediction 
                      ($\alpha_s\!=0.116,\:\delta_\alpha\!=0$)}\\
 \hline
         $m_t$ (GeV)   &               
 &    150 &    150 &    175 &    175\\
         $m_H^{}$ (GeV)&               
 &    100 &   1000 &    100 &   1000\\
 \hline
\multicolumn{1}{|l|}{$\nu$--$q$\cite{hhkm,nuq_dat}
\hphantom{\hspace{2cm}}} 
& & & & &\\
$g_L^2$         &  0.2980 $\pm$ 0.0044 
 & 0.2976 & 0.2955 & 0.2995 & 0.2973\\
$g_R^2$         &  0.0307 $\pm$ 0.0047 
 & 0.0296 & 0.0298 & 0.0295 & 0.0297\\
$\delta_L^2$    & -0.0589 $\pm$ 0.0237 
 &-0.0633 &-0.0632 &-0.0634 &-0.0634\\
$\delta_R^2$    &  0.0206 $\pm$ 0.0160 
 & 0.0177 & 0.0178 & 0.0177 & 0.0178\\
$\chi^2/({\rm d.o.f.}) $       &                      
 &   0.21/4 &   0.77/4 &   0.25/4 &   0.25/4\\[2mm]
\multicolumn{1}{|l|}{$\nu$--$e$\cite{beyer}} & & & & & \\
$s^2_{eff}$     &   0.233 $\pm$  0.008 
 &  0.231 &  0.232 &  0.230 &  0.231\\
$\rho_{eff}$    &   1.007 $\pm$  0.028 
 &  1.011 &  1.009 &  1.013 &  1.011\\
$\chi^2/({\rm d.o.f.})$        &                      
 &   0.09/2 &   0.02/2 &   0.18/2 &   0.06/2\\[2mm]
\multicolumn{1}{|l|}{APV\cite{apv_dat}} & & & & & \\
$Q_W$           &  -71.04 $\pm$   1.81 
 & -73.20 & -73.30 & -73.20 & -73.30\\
$\chi^2/({\rm d.o.f.})$        &                      
 &   1.42/1 &   1.56/1 &   1.43/1 &   1.57/1\\[2mm]
\multicolumn{1}{|l|}{$e$--D\cite{ed_dat,hhkm}} & & & & & \\
$2C_{1u}-C_{1d}$&   0.938 $\pm$  0.264 
 &  0.719 &  0.713 &  0.723 &  0.717\\
$2C_{2u}-C_{2d}$&  -0.659 $\pm$  1.228 
 &  0.099 &  0.092 &  0.104 &  0.096\\
$\chi^2/({\rm d.o.f.})$        &                      
 &   1.43/2 &   1.69/2 &   1.27/2 &   1.51/2\\[2mm]
 \hline
\multicolumn{1}{|l|}{$W$ mass\cite{pdg94,mw93}} & & & & & \\
$m_W$           &   80.24 $\pm$   0.16 
 &  80.25 &  80.08 &  80.40 &  80.23\\
$\chi^2/({\rm d.o.f.})$        &                      
 &   0.00/1 &   0.96/1 &   1.04/1 &   0.00/1\\
 \hline
 \end{tabular}
\end{center}
}
\def\tableofstunew{%
\begin{center}
\begin{tabular}{cccccc}
   \hline
   \begin{tabular}{cc} $m_t$\\(GeV) \end{tabular}
   & 
   \begin{tabular}{cc} $\mh$\\(GeV) \end{tabular}
   & 
   $\displaystyle{\left(
   \begin{array}{c} S \\ T \\ U \end{array}
   \right)}$
   &
   $\displaystyle{\left(
   \begin{array}{c}S_{\rm SM}\\T_{\rm SM}\\U_{\rm SM}\end{array}
   \right)}$
   &
   $\displaystyle{\left(
   \begin{array}{c}S_{\rm new}\\T_{\rm new}\\U_{\rm new}\end{array}
   \right)}$
   & 
   $\chi^2_{\rm min}/({\rm d.o.f.})$ \\
   \hline
   150 & 100 &
   $\displaystyle{\begin{array}{r} 
    -0.22\pm 0.20 \\0.84\pm 0.20 \\-0.05\pm 0.50
   \end{array}}$
   &
   $\displaystyle{\begin{array}{r} 
    -0.21\\0.59\\0.30
   \end{array}}$
   &
   $\displaystyle{\begin{array}{r} 
   -0.01\pm 0.20 \\0.25\pm 0.20 \\-0.35\pm 0.50
   \end{array}}$
   &
    17.5/18
   \\ 
   \hline
   150 & 1000 &
   $\displaystyle{\begin{array}{r} 
      -0.23\pm 0.20 \\ 0.83\pm 0.20\\ -0.05\pm 0.50
   \end{array}}$
   &
   $\displaystyle{\begin{array}{r} 
      -0.06\\  0.30\\  0.29
   \end{array}}$
   &
   $\displaystyle{\begin{array}{r} 
     -0.17\pm 0.20 \\  0.53\pm 0.20 \\ -0.34\pm 0.50
   \end{array}}$
   &
    17.6/18
   \\
   \hline
   175 & 100 &
   $\displaystyle{\begin{array}{r} 
      -0.18\pm  0.20\\ 0.92\pm  0.20\\ -0.12\pm  0.50
   \end{array}}$
   &
   $\displaystyle{\begin{array}{r} 
      -0.23\\  0.89\\  0.36
   \end{array}}$
   &
   $\displaystyle{\begin{array}{r} 
      0.05\pm 0.20 \\  0.03\pm 0.20 \\ -0.48\pm 0.50
   \end{array}}$
   &
   20.6/18
   \\
   \hline
   175 & 1000 &
   $\displaystyle{\begin{array}{r} 
      -0.19\pm  0.20\\ 0.91\pm  0.20\\ -0.12\pm  0.50
   \end{array}}$
   &
   $\displaystyle{\begin{array}{r} 
       -0.08\\  0.59\\  0.35
   \end{array}}$
   &
   $\displaystyle{\begin{array}{r} 
        -0.12\pm 0.20 \\  0.32\pm 0.20 \\ -0.47\pm 0.50
   \end{array}}$
   &
   20.7/18
   \\
   \hline
\end{tabular}
\end{center}
}

\def\tcaptionofdata{%
Experimental data on $Z$-pole and the SM predictions. 
}

\def\tcaptionofsbar{%
$\protect\sbar^2(\protect\mmz)$ from various asymmetries.
The column `$\chi$' denotes the deviation from the combined 
mean value normalized by each error.
}
 
\def\tcaptionofdelb{%
$\protect\delb(\mmz)$ as determined from various measurements. 
The dependences on the other parameters are shown 
by the coefficients $C({\protect\gzbar^2})$, 
$C({\protect\sbar^2})$ and $C({\alpha_s})$. 
See eq.~(\protect\ref{parametrization_of_delb}). 
}

\def\tcaptionofalps{%
$\alpha_s\equiv\alpha_s(\mz)_{\msbar}$ determined from the 
electroweak data by assuming the SM. The SM predictions for 
$\gzbar^2(\mmz)$, $\sbar^2(\mmz)$ and $\delb(\mmz)$ are also shown. 
}

\def\tcaptionofotherdata{%
Data from low-energy neutral-current experiments and 
$W$-mass measurements that are used in our analysis. 
The SM prediction is also shown.
}

\def\tcaptionoflencatmz{%
$\protect\gzbar^2(\mmz)$ and $\protect\sbar^2(\mmz)$ 
determined from low-energy neutral-current experiments. 
The running of the charge form-factor are calculated 
in the SM with $m_t=175\gev$ and $\mh=100\gev$. 
}

\def\tcaptionofstunew{%
Constraints on the parameters 
$S_{\rm new}$, $T_{\rm new}$, $U_{\rm new}$ 
which are obtained by subtracting the SM contribution 
$S_{\rm SM}$, $T_{\rm SM}$, $U_{\rm SM}$ from 
$S$, $T$, $U$ for $\alpha_s=0.116$ and 
$\delta_\alpha=0$. 
Correlations among errors are the same as in 
eq.~(\protect\ref{fitofstu_mean}). 
}

\def\fcaptionofgzbsb{%
A two-parameter fit to the $Z$ boson parameters in
the ($\protect\sbar^2(\mmz), \protect\gzbar^2(\protect\mmz)$) plane. 
The $\protect\zbb$ vertex form-factor, 
$\protect\delb(\protect\mmz)$, 
and the QCD coupling, $\alpha_s(\protect\mz)$, 
are treated as external parameters. 
The 1-$\sigma$ contours are shown for two values of 
$\alpha_s(\protect\mz)$, 
0.116 (solid lines) and 0.124 (dashed lines), 
and for two values of $\protect\delb(\protect\mmz)$, 
$-0.0100$ (thick lines) and $-0.0079$ (thin lines). 
In the SM, $\delb=-0.0100$ corresponds to $m_t=175\protect\gev$,  
while $\delb=-0.0079$ corresponds to $m_t=150\protect\gev$. 
Also shown are the SM predictions in the range 
$125\protect\gev<m_t<225\protect\gev$ and 
$10\protect\gev<\protect\protect\mh<1000\protect\gev$, 
which are obtained by assuming $\alpha_s(\protect\mz)=0.116$ and 
$\delta_\alpha \equiv 1/\bar{\alpha}(\protect\mmz)-128.72 =0$.
}

\def\fcaptionofstu{%
A two-parameter fit to the $Z$ boson parameters in the 
($S$, $T$) plane for $\delta_\alpha=0$ and $\protect\delg=0.0055$. 
The $\protect\zbb$ vertex form-factor, 
$\protect\delb(\protect\mmz)$, and the QCD coupling, $\alpha_s(\mz)$, 
are treated as external parameters in the fit. 
The 1-$\sigma$ contours are shown for 
two values of $\alpha_s$, 0.116 (solid lines) and 0.124 (dashed lines), 
and for two values of $\protect\delb$, 
$-0.0100$ (thick lines) and $-0.0079$ (thin lines). 
In the SM, $\delb=-0.0100$ corresponds to $m_t=175\protect\gev$,  
while $\delb=-0.0079$ corresponds to $m_t=150\protect\gev$. 
Also shown are the SM predictions in the range 
$125\protect\gev\!<\!m_t\!<\!225\protect\gev$ 
and $50\protect\gev\!<\!\protect\mh\!<\!1000\protect\gev$. 
The mild dependence on $m_t$ and $\mh$ 
in the SM running between $q^2=0$ and $q^2=\protect\mmz$ is 
calculated with $\protect\mh=100\protect\gev$ and the above 
values of $m_t$ determined from $\protect\delb(\protect\mmz)$. 
The estimates\protect\cite{stu} for one doublet 
${\protect\rm SU}(N_c)$--TC models are shown for $N_c=2,3,4$. 
}

\def\fcaptionofmtmh{%
The SM fit to all electroweak data in the ($\protect\mh,\,m_t$) 
plane for (a) $\alpha_s=0.116$ and (b) 0.124, 
with $\delta_\alpha=0$. 
The thick inner and outer contours correspond to 
$\Delta\chi^2=1$ ($\sim$ 39\% CL), 
and $\Delta\chi^2=4.61$ ($\sim$ 90\,\%~CL), respectively. 
The minimum of $\chi^2$ is marked by an ``$\times$''. 
Also shown are the 1-$\sigma$ constraints from the 
$Z$-pole asymmetries, $\Gamma_Z$ and $\mw$. 
The dashed lines show the constraint only from 
$R_\ell$ and $R_b$ (see also Fig.~\protect\ref{figureofrlrb}). 
They correspond to 
$\chi^2=2.25,\,4.0,\,6.25,\,9.0$. 
The regions $m_t<131\gev$ and $\protect\mh<63\gev$ are excluded by 
CDF\protect\cite{top_d0} and LEP experiments\protect\cite{mh_limit},
respectively.
}
 
\def\fcaptionofrlrb{%
The $R_b$ vs. $R_\ell$ plane. 
The SM predictions are shown in the range 
$120\protect\gev<m_t<240\protect\gev$, 
and $60\protect\gev<\protect\mh<1000\protect\gev$, 
for two cases of $\alpha_s$ ($\alpha_s=0.116$ and $0.124$).
Also shown are the $\chi^2=1,4,9$ contours obtained by combining 
the $R_\ell$ and $R_b$ data. 
}

\def\fcaptionofmtmhda{%
The SM fit to all electroweak data in the ($\protect\mh,\,m_t$) 
plane for $\alpha_s=0.116$. 
The dashed lines show the case for 
$\delta_\alpha=-0.1$, while 
the solid lines show for $\delta_\alpha=+0.1$.
The inner and outer contours correspond to 
$\Delta\chi^2=1$ and $\Delta\chi^2=4.61$, respectively. 
}

\def\fcaptionofmhcl{%
Constraints on the Higgs mass in the SM from all the 
electroweak data. 
Upper and lower bounds of the Higgs mass at 95\% CL are 
shown as functions of the top mass $m_t$, where 
$m_t$ is treated as an external parameter
with negligible uncertainty.
The thick solid lines show the case for $\alpha_s=0.116$ 
and $\delta_\alpha\equiv 1/\bar{\alpha}(\mmz)-128.72=0$. 
The cases for $\alpha_s=0.111$ and $0.121$ with $\delta_\alpha=0$ 
are shown by thick dashed and dot-dashed lines. 
The cases for of $\delta_\alpha=\pm 0.1$ with $\alpha_s=0.116$ 
are also shown by thin dashed and dot-dashed lines. 
}

\begin{document}
\thispagestyle{empty}
\vglue 10mm
\vspace*{-10mm}
\baselineskip10pt
\begin{flushright}
\begin{tabular}{l}
{\bf KEK-TH-418}\\
{\bf KEK preprint 94--147}\\
{\bf KANAZAWA-94-24}\\
November 1994
\end{tabular}
\end{flushright}
\baselineskip18pt 
\vglue 20mm 
\begin{center}{\large \bf
Constraints on the electroweak universal parameters and \\
the top and Higgs masses from updated LEP/SLC data
}\end{center}
\vglue 10mm
\def\thefootnote{\fnsymbol{footnote}}
\setcounter{footnote}{0}
\begin{center}
{\bf S.~Matsumoto}\\[4mm]
{\it Theory Group, KEK, Tsukuba, Ibaraki 305, Japan}\\[2mm]
{and}\\[2mm]
{\it Department of Physics, Kanazawa University, 
     Kanazawa, 920-11, Japan}
\end{center}
\vspace{10mm}
\begin{center}
{\bf ABSTRACT}
\\[6mm]
\begin{minipage}{14cm}
{
A global analysis is performed using the latest data from LEP and SLC. 
Constraints on the electroweak universal parameters ($S,T,U$) and on 
the masses of the top quark and Higgs boson within the Standard 
Model (SM) are investigated. The uncertainties due to the QCD and QED 
effective couplings, $\alpha_s(\mz)$ and $\bar{\alpha}(\mmz)$, are 
examined in detail. Even though the mean value of $S$ is increased to 
be consistent with zero, the naive Technicolor models are still 
disfavored due to its reduced error. Within the SM, we find the 
90\,\%CL constraints; $133\gev<m_t<190\gev$ and $10\gev<\mh<440\gev$ 
for $\alpha_s(\mz)=0.116$ and $1/\bar{\alpha}(\mmz)=128.72$. 
The experimental constraints on the $\zbb$ vertex 
form-factor, $\delb(\mmz)$, play an important role in disfavoring 
the region of large $m_t(m_t \sim 200\gev)$ and large 
$\mh(\mh \sim 1000\gev)$. If $m_t$ is precisely known, the present 
electroweak data give a rather strict upper bound on the Higgs mass, 
$\mh<140\,(300)\gev\,$ at 95\%~CL, for $m_t=160\,(175)\gev$ 
and for the above $\alpha_s(\mz)$ and $\bar{\alpha}(\mmz)$. 
}
\end{minipage}
\end{center}

\vglue 15mm
\baselineskip18pt  

\newpage
\def\thefootnote{\fnsymbol{footnote}}
\setcounter{footnote}{0}

During 1993 the four LEP experiments performed a high precision 
scan near the $Z$ boson resonance\cite{lep94,lep94_glasgow}. 
The uncertainties in the $Z$ parameters, 
such as the total $Z$ width and the various asymmetries, 
are significantly reduced from the previous results\cite{lep93}. 
Also much improved is the measurement of the left-right 
polarization asymmetry at SLC\cite{alr94}. 
Additionally, the $W$-mass measurements at Tevatron were 
also improved in 1993\cite{mw93}. 
More striking is evidence for the top quark 
reported by the CDF 
Collaboration\cite{top_cdf}. 
These data may provide hints about new physics beyond the 
Standard Model (SM) through quantum effects 
prior to its discovery at future collider experiments. 

Hence it is important to interpret various data in a systematic 
way that is convenient not only for testing the SM, 
but also for studying consequences of new physics. 
In this letter we present an update of the comprehensive study of 
the electroweak data based on the formalism of ref.~\cite{hhkm}. 
A theoretical fit of the electroweak data has been performed 
by allowing the gauge boson propagator corrections 
and the $\zbb$ vertex form-factor to vary freely within the 
generic ${\rm SU(2)_L\times U(1)_Y}$ gauge theory framework; 
SM dominance of the remaining vertex and box corrections 
has been assumed. 
The formalism allows us to obtain constraints on the universal 
$(S,T,U)$ parameters\cite{stu} which are modified 
in ref.~\cite{hhkm} to include the SM contribution, 
as functions of $\alpha_s(\mz)$, 
the QED effective charge, $\bar{\alpha}(\mmz)$, 
and the $\zbb$ vertex correction factor, 
$\delb(\mmz)$, which are not precisely known at present. 
By neglecting new physics contributions to the parameters 
$S,T,U$ and $\delb(\mmz)$, we obtain constraints on $m_t$ and $\mh$ 
as functions of $\alpha_s$ and $\bar{\alpha}(\mmz)$. 
All results are expressed such that consequences of 
future improvements in the estimate of $\alpha_s(\mz)$ and 
$\bar{\alpha}(\mmz)$ are immediately transparent. 

We start with a brief review, but for details and further 
references the reader is referred to ref.~\cite{hhkm}. 
Then we discuss the significance of the updated data on the 
universal parameters of the neutral-current sector, 
($\gzbar^2(\mmz), \sbar^2(\mmz)$), which are nearly 
equivalent to the modified ($S,T$) parameters of ref.~\cite{hhkm}. 
Determination of the $\zbb$ vertex correction factor $\delb(\mmz)$ 
and that of $\alpha_s$ from various electroweak measurements 
are also discussed. 
We also discuss the situation where all radiative effects are 
dominated by the SM contribution; here we place constraints 
on the masses of the top quark and the Higgs boson 
which are the only relevant free parameters of the theory. 
Finally, we modify this discussion to the scenario where 
$m_t$ is measured accurately. 

In generic ${\rm SU(2)_L \times U(1)_Y}$ theories, 
the universal effective form-factors 
that characterize the gauge boson propagator corrections 
are given by
  \begin{subequations}
   \label{rg}
  \begin{eqnarray}
     \frac{1}{\ebar^2(q^2)} &=& \frac{1}{\ehat^2(\mu)}
     \Bigl[\,1 +{\rm Re}\pibar_{T,\gamma}^{\gamma\gamma}(q^2)\,\Bigr]\,,
    \label{ebar}
  \\
     \sbar^2(q^2) &=& \shat^2(\mu)
      +\frac{\ebar^2(q^2)}{\ehat(\mu)\gzhat^{}(\mu)}
     \,{\rm Re}\,\pibar_{T,\gamma}^{\gamma Z}(q^2),
    \label{sbar}
  \\
     \frac{1}{\gzbar^2(q^2)} &=& \frac{1}{\gzhat^2(\mu)}
     \Bigl[\,1 +{\rm Re}\pibar_{T,Z}^{ZZ}(q^2)\,\Bigr]\,,
    \label{gzbar}
  \\
     \frac{1}{\gwbar^2(q^2)} &=& \frac{1}{\ghat^2(\mu)}
     \Bigl[\,1 +{\rm Re}\pibar_{T,W}^{WW}(q^2)\,\Bigr]\,,
    \label{gwbar}
  \end{eqnarray}
  \end{subequations}
where the hatted couplings, 
$\ehat \equiv \ghat\shat \equiv \gzhat^{}\shat\chat$, 
and all ultraviolet-singular loop functions are renormalized 
in the $\msbar$ scheme. 
We also use the notation 
$
     \pibar_{T,V}^{AB}(q^2)
      \equiv [\pibar_T^{AB}(q^2) -\pibar_T^{AB}(\mmv)]/(q^2 -\mmv)
$;
$\mv$ is the physical mass of the gauge boson `$V$' 
(that is, $\mv=\mw$, $\mz$ or $m_{\gamma}$ with $m_\gamma =0$) 
and the subscript $T$ denotes the transverse part of the 
vacuum polarization tensor, $\Pi_{\mu \nu}(q)$.  
The `overlines' denote inclusion of the pinch terms\cite{kl,pt2,pt3}. 
The explicit expressions for $\pibar$'s in the SM 
are found in ref.~\cite{hhkm}. 
The helicity amplitudes of neutral-current processes 
are expressed in terms of these charge form-factors 
plus appropriate vertex and box corrections. 
Hence the charge form-factors 
can be directly extracted from the experimental data  
by assuming SM dominance to the vertex and box corrections, 
and the extracted values can be compared with various 
theoretical predictions. 

\begin{table}[t]
\caption{\tcaptionofdata}
\label{tableofdata}\vspace{-2mm}
\tableofdata
\end{table}
%
The experimental data on the $Z$-pole\cite{lep94_glasgow,alr94} which 
are used in our analysis are listed in Table~\ref{tableofdata}. 
Also shown are the SM predictions for 
$(m_t,\,\mh)=(150\gev,\,100\gev)$, $(150\gev,\,1000\gev)$, 
$(175\gev,\,100\gev)$ and $(175\gev,\,1000\gev)$, 
with $\alpha_s(\mz)=0.116$\cite{pdg94}%
\footnote{
 \normalsize \baselineskip 18pt
  {}From the various measurements of $\alpha_s(\mz)$ as summarized 
  by the Particle Data Group (PDG)\cite{pdg94}, 
  we obtain $\alpha_s(\mz)_{\msbar}=0.116\pm 0.0024$ 
  (statistical error only) by excluding the data from the $Z$ 
  parameter that will be discussed separately in this letter. 
  The PDG assign an error of $\pm 0.005$ to account for the 
  theoretical uncertainties. 
 }
and $\delta_\alpha\equiv 1/\bar{\alpha}(\mmz) -128.72=0$\cite{hhkm}, 
where $\bar{\alpha}(\mmz)\equiv\ebar^2(\mmz)/4\,\pi$. 
The correlations in the errors of the $Z$ line-shape parameters 
($\mz,\Gamma_Z,\sigma_h,R_\ell,A_{\rm FB}^{0,\ell}$) 
and those of the measurements concerning the bottom and charm quarks 
($R_b,R_c,A_{\rm FB}^{0,b},A_{\rm FB}^{0,c}$) 
as reported in ref.~\cite{lep94} are taken into account in the fits. 
The $Z$ mass, $\mz\!=\!91.1888\gev$, is treated as an input 
parameter neglecting its error. 
This is justified because the experimental uncertainty and 
correlations are so small. 
In the following analysis, 
(a) we assume that only three neutrinos ($N_\nu = 3$) contribute 
    to the invisible width of $Z$, 
(b) we include perturbative QCD corrections with finite 
    quark-mass effects, 
(c) we calculate vertex and box corrections within the SM, 
(d) except for corrections due to the $\zbb$ vertex. 
\label{assumption}
The $\zbb$ vertex correction is represented by the form-factor 
$\delb(\mmz)$ which, in many cases, is treated as an external 
parameter similar to the treatment of $\alpha_s(\mz)$. 
The SM predicts $\delb=-0.0079, -0.0100, -0.0123$ for 
$m_t=150,175,200\gev$, respectively, 
with $\mh=100\gev$ and $\alpha_s=0.116$;
we make the abbreviations $\delb\equiv\delb(\mmz)$ and 
$\alpha_s\equiv\alpha_s(\mz)_{\msbar}$. 
Dependence on $\mh$ and $\alpha_s$ at the two-loop level 
is present but is not significant\cite{hhkm}.

Under these conditions all $Z$ parameters are expressed 
in terms of $\gzbar^2(\mmz)$, $\sbar^2(\mmz)$, $\delb(\mmz)$ 
and $\alpha_s(\mz)$.  
Among these form-factors $\sbar^2(\mmz)$ is determined primarily 
from the asymmetry measurements. 
%
\begin{table}[t]
 \caption{\tcaptionofsbar}
 \label{tableofsbar}\vspace{-2mm}
  \tableofsbar
\end{table}
%
Table~\ref{tableofsbar} shows the fitted values of $\sbar^2(\mmz)$ 
as determined from each asymmetry measurement. 
The dependence of the fits on the remaining parameters, 
$\gzbar^2(\mmz)$, $\delb$ and $\alpha_s$, 
is negligible compared to the errors. 
Also shown are two combined fits. 
One includes the leptonic asymmetries only and the other includes 
all the asymmetry measurements. 
For the combined fits $\chi^2_{\rm min}$ per degree of 
freedom is given. 
The deviation of each individual fit from the latter combined fit 
is given in the column $\chi$, defined as 
%
$
    \chi \equiv \frac{\langle\sbar^2(\mmz)\rangle -0.2306}
                     {\Delta\sbar^2(\mmz)}\,,
$
%
where $\langle\sbar^2(\mmz)\rangle$ and $\Delta\sbar^2(\mmz)$ denote, 
respectively, the mean value and the error of $\sbar^2(\mmz)$ 
as determined from each data. 
The total $\chi^2/({\rm d.o.f.})$ is 
$9.4/5$ corresponding to 9\% confidence level; 
this reflects the fact that the left-right asymmetry data 
from SLC gives a somewhat smaller value of $\sbar^2(\mmz)$ than 
the other data. 

The only quantity which is sensitive to $\gzbar^2(\mmz)$ 
is the total width of the $Z$ boson, $\Gamma_Z$, which also depends 
on $\sbar^2(\mmz)$, $\delb$ and $\alpha_s$. 
On the other hand, $R_\ell$ and $\sigma_h^0$ depend 
on $\sbar^2(\mmz)$ and $\delb$, 
and they depend strongly on $\alpha_s$. 
The ratios $R_b$ and $R_c$ are sensitive only to $\delb$, 
that is, they hardly contribute to the universal parameter fits, 
but have little dependence upon $\alpha_s$. 
By taking $\gzbar^2(\mmz)$ and $\sbar^2(\mmz)$ as fit variables 
while treating $\delb$ and $\alpha_s$ as external variables, 
we obtain from the two-parameter fit to all the $Z$ parameters~: 
%
\fitofgzbsb
%
where the errors and the correlation are almost 
independent of $\alpha_s$ and $\delb$. 
It should be noted that 
the quantities which are sensitive to $\alpha_s$ 
($\Gamma_Z$, $\sigma_h^0$ and $R_\ell$) 
are also sensitive to $\delb$, 
but they depend on the two parameters only through the combination 
 \begin{eqnarray}
    \alpha_s' \equiv \alpha_s(\mz) +1.6\,\delb(\mmz)\,. 
    \label{delb_alps}
 \end{eqnarray}
This is because the hadronic contributions to these quantities 
arise from just one quantity, $\Gamma_h$, 
which depends on $\alpha_s$ and $\delb$ in 
in approximately the above combination\cite{hhkm}. 
%

\begin{figure}[b]
\begin{center}
\leavevmode\psfig{file=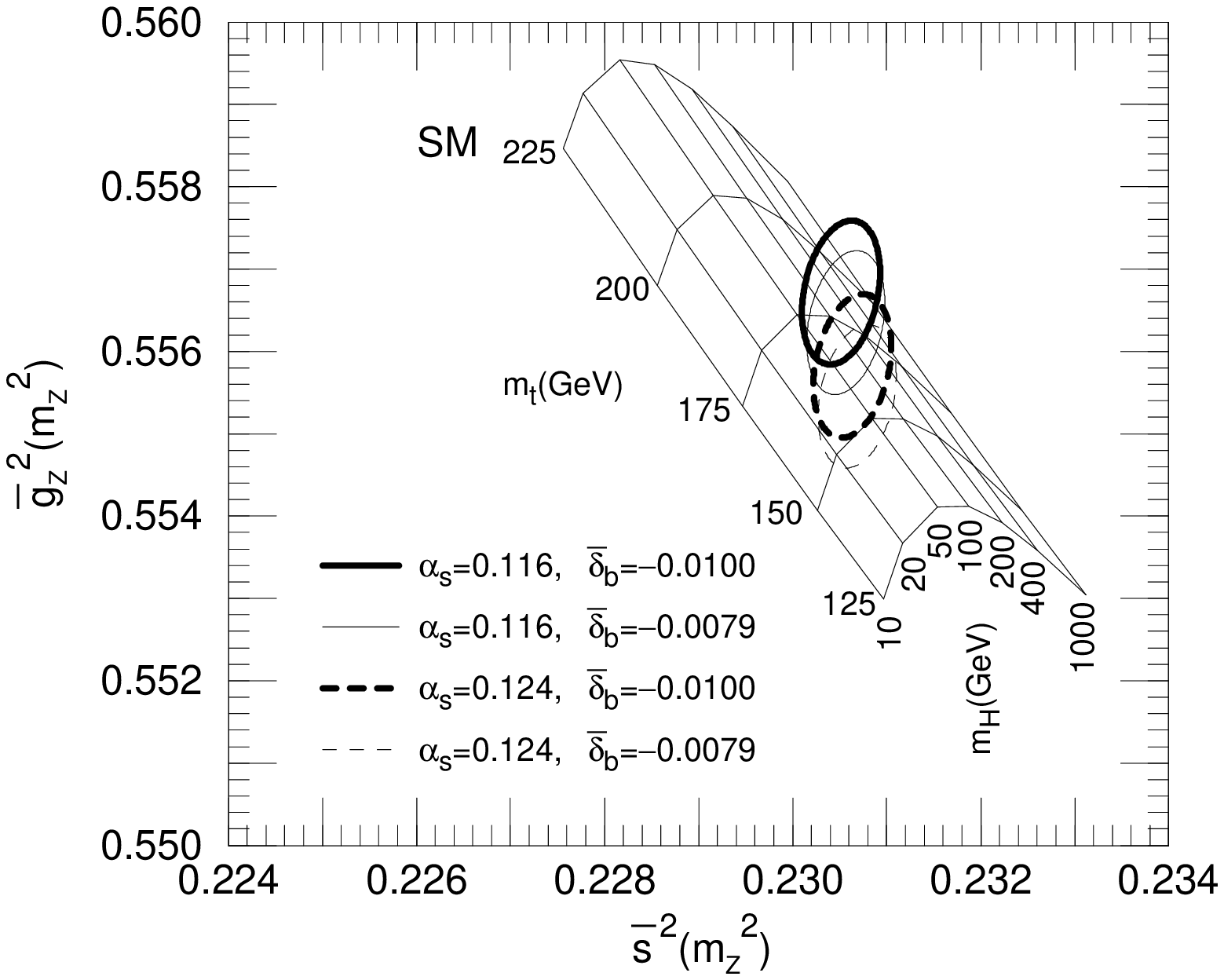,width=12cm,silent=0}
\end{center}\vspace{-7mm}
\caption{\fcaptionofgzbsb}
\label{figureofgzbsb}
\end{figure}

The results of the fit (\ref{fitofgzbsb}) are displayed in 
Fig.~\ref{figureofgzbsb}. 
We present 1-$\sigma\;$ allowed contours 
in the ($\sbar^2(\mmz)\,,\gzbar^2(\mmz)$) plane for 
$\alpha_s=0.116$ (solid lines) and $0.124$ (dashed lines),  
and for $\delb=-0.0100$ (thick lines) and $-0.0079$ (thin lines). 
Also shown by the lattices are the SM predictions for 
$125\gev<m_t<225\gev$ and $50\gev<\mh<1000\gev$. 
In these SM predictions all known two-loop 
corrections of the $\O(m_t^4)$ and $\O(\alpha\alpha_s)$ 
level 
have been included 
(see ref.~\cite{hhkm} for references). 
We set $\alpha_s=0.116$ and $\delta_\alpha=0$ 
when calculating the SM prediction. 
In the SM the present estimate\cite{piqq_h_latest} of the hadronic 
contribution to the running of the charge form-factor, 
$\bar{\alpha}(q^2)$, is 
$\delta_\alpha = 0\pm 0.10$\cite{hhkm}. 
While changing $\alpha_s$ by $\pm 0.005$\cite{pdg94} 
has little effect, changing $\delta_\alpha$ by $\pm$0.10 
leads to a shift in the SM prediction for $\sbar^2(\mmz)$ by 
$\mp 0.00026$; this is more than half of its uncertainty. 

So far we have treated $\delb$ and $\alpha_s$ as 
external parameters. 
However, in principle, they can be extracted from the data 
once the remaining parameters are determined 
by other measurements, 
or they may be calculated in a specific theoretical model. 

The parameter $\delb$ is determined from the ratio $R_b$ 
almost independently of the other parameters 
($\gzbar^2(\mmz)$, $\sbar^2(\mmz)$ and $\alpha_s$), 
while its extraction from $\Gamma_Z$, $\sigma_h^0$ and $R_\ell$ 
depends more heavily on these parameters. 
%
\begin{table}[t]
\caption{\tcaptionofdelb}
\label{tableofdelb}\vspace{-2mm}
\tableofdelb
\end{table}
%
We may parametrize $\delb$ obtained from each measurement 
in terms of $\gzbar^2(\mmz)$, $\sbar^2(\mmz)$ and $\alpha_s$ 
as follows~:
 \begin{eqnarray} 
    \label{parametrization_of_delb}
    \delb &=&
    \langle \delb \rangle +C(\gzbar^2)\bydgzb +C(\sbar^2)\bydsb 
          +C(\alpha_s)\bydalps \:  \pm \Delta\delb,
 \end{eqnarray}
where $\langle \delb \rangle$ and $\Delta\delb$ denote the mean 
value and the error of the individual measurement, respectively, 
while the coefficients $C(\gzbar^2)$, $C(\sbar^2)$ and 
$C(\alpha_s)$ show its dependence on the remaining parameters. 
In Table~\ref{tableofdelb} we show the results 
when $\delb$ is extracted from 
$\Gamma_Z$, $\sigma_h^0$, $R_\ell$ and $R_b$.
The same coefficient, $C(\alpha_s)\approx -0.0031$, 
appears for $\Gamma_Z$, $\sigma_h^0$, $R_\ell$, 
as a consequence of their dependence on the combination 
$\alpha_s'$ (\ref{delb_alps}). 
The present data on $R_b$ gives a value of $\delb(\mmz)$ 
which is about 2 standard deviations larger than 
the SM prediction for $m_t\sim 175\gev$ ($\delb\sim -0.0100)$. 
It is further noted that, for $\alpha_s = 0.116$, 
the data on $\Gamma_Z$ and $R_\ell$ also give somewhat larger 
values of $\delb$ than the SM prediction. 
Therefore we have to assume larger value of $\alpha_s$ 
($\alpha_s \sim 0.125$) 
in order to get values of $\delb$ consistent with the SM 
(See Table~\ref{tableofalps}). 
Refering back to Table~\ref{tableofdelb}, we give 
in the bottom line 
$\delb$ as determined from all measurements of $Z$ parameters, 
that is, including asymmetries and $R_c$ 
together with the error correlations. 
It is clear that $\delb$ is primarily determined from 
$\Gamma_Z$ and $R_\ell$, and hence it depends strongly on $\alpha_s$. 
An accurate measurement of $R_b$ offers the key to 
separate the measurements of $\alpha_s$ and $\delb$. 

Similarly, $\alpha_s$ can be extracted 
from the electroweak data alone 
once the form-factors $\gzbar^2(\mmz)$, $\sbar^2(\mmz)$ and 
$\delb(\mmz)$ are given by a specific model. 
From the global fit we find 
%
\begin{table}[t]
\caption{\tcaptionofalps}
\label{tableofalps}\vspace{-2mm}
\tableofalps
\end{table}
%
\fitofalps
%
Here the reference values, 
$\gzbar^2(\mmz)\!=\!0.55550$, $\sbar^2(\mmz)\!=\!0.23068$ and 
$\delb\!\equiv\!\delb(\mmz)\!=\!-0.0034$, are the best-fit values 
when we make a three-parameter fit in terms of 
($\gzbar^2(\mmz),\sbar^2(\mmz),\delb$). 
The quantities that are most sensitive to $\alpha_s$ are 
$\Gamma_Z$ and $R_\ell$. 
In fact, one can obtain the above results (\ref{fitofalps}) 
by solving eq.~(\ref{parametrization_of_delb}) 
for $\Gamma_Z$ and $R_\ell$ 
using the values in Table~\ref{tableofdelb}. 
In the SM the form-factors $\gzbar^2(\mmz)$, $\sbar^2(\mmz)$ 
and $\delb(\mmz)$ are functions of $m_t$ and $\mh$, 
and hence 
$\alpha_s$ can be determined as a function of $m_t$ and $\mh$.  
Table~\ref{tableofalps} shows the extracted values of $\alpha_s$ 
for several set of $m_t$ and $\mh$ with $\delta_\alpha=0$, 
together with the SM prediction for the three form-factors. 
For the given sets of $m_t$ and $\mh$ in the table 
the extracted value of $\alpha_s$ is somewhat larger than the 
estimate of the PDG\cite{pdg94}$^*$ 
which is $\alpha_s=0.116\pm 0.005$. 

If we allow all the four parameters, $\gzbar^2(\mmz)$, 
$\sbar^2(\mmz)$, $\delb(\mmz)$ and $\alpha_s$, 
to be fitted by the data we find
  \begin{subequations}
  \begin{eqnarray}
    &&
    \left.
    \begin{array}{c@{\!\;}r@{\!\;}r@{\!\;}l}
      \gzbar^2(\mmz) &=& 0.55570 &\pm  0.00108 \\
      \sbar^2(\mmz)  &=& 0.23066 &\pm  0.00043 \\
      \delb(\mmz)    &=& -0.0017 &\pm  0.0049  \\
      \alpha_s       &=& 0.1117  &\pm  0.0093 
    \end{array}
    \right\} 
    \quad \rho_{\rm corrr} = 
    \left(
    \begin{array}{rrrr}
        1.00 & 0.10 &  0.01 & -0.36 \\
             & 1.00 &  0.01 &  0.12 \\
             &      &  1.00 & -0.80 \\
             &      &       &  1.00 
    \end{array}
    \right)\,,
    \\
    && \quad \chi^2_{\rm min} =  11.4. 
  \end{eqnarray}
  \end{subequations}
This is, of course, consistent with the 
parametrization (\ref{fitofgzbsb}). 
The rather small mean value for $\alpha_s$ is a consequence of 
a large $\delb$ which is prefered by the data on $R_b$, 
as explained above 
(see eq.(\ref{fitofgzbsbchisq}) and Table~\ref{tableofdelb}).

It is often important to obtain the constraint on the number 
of neutrinos, $N_\nu$, which, in the above analysis, 
has been assumed to be $N_\nu=3$. 
We consider here an analysis without the condition, 
$N_\nu=3$.
By treating $N_\nu$, $\gzbar^2(\mmz)$ and $\sbar^2(\mmz)$ 
as the three parameters of our fit 
we obtain~:
%
\fitofgzbsbnnu
%
This result strongly supports the validity of the assumption~(a) 
in page \pageref{assumption}. 
We can also find the best-fit value of $N_\nu$ 
from the above result as functions of 
$\gzbar^2(\mmz)$, $\sbar^2(\mmz)$, $\delb(\mmz)$ and $\alpha_s$~:
%
\fitofnnu
%
where the reference values for the three form-factors 
are chosen as in eq.~(\ref{fitofalps}). 


The above results may be re-interpreted in the language 
of $S$, $T$ and $U$\cite{hhkm,stu}. 
When the new physics scale is higher than the scale 
of precision measurements new-physics contributions 
to the running of the charge form-factors may be neglected. 
In such a case one may combine the low-energy neutral-current 
experiments which determine the form-factors 
$\gzbar^2(0)$ and $\sbar^2(0)$\cite{hhkm} 
with the $Z$ parameter measurements. 
Here one assumes that the running of $\gzbar^2(q^2)$ 
and $\sbar^2(q^2)$ between $q^2=0$ and $q^2=\mmz$ is governed 
only by SM physics. 
The universal propagator corrections 
in the neutral-current sector are then parametrized 
by just two parameters, essentially $S$ and $T$. 
The $U$ parameter is determined from the 
charged-current sector through the charge form-factor 
$\gwbar^2(0)$\cite{hhkm} which is determined from 
measurements of the $W$-boson mass using the relation
 \begin{eqnarray}
   G_F &=& \frac{\gwbar^2(0)+\ghat^2\delg}{4\,\sqrt{2}\mmw} \,.
   \label{gf}
 \end{eqnarray} 
Here $\delg$ is the vertex and box correction to the muon 
lifetime\cite{del_gf} after subtraction of the pinch term. 
In the SM, $\delg=0.0055$\cite{hhkm}. 
We adopt a modified version of the original $S$, $T$ and $U$ 
parameters which includes the SM radiative effects as well 
as new physics contributions\cite{pt3,hhkm}. 
They are related to the charge form-factors by the following 
identities\cite{hhkm}: 
  \begin{subequations}
    \label{gbarfromstu}
  \begin{eqnarray}
     \frac{1}{\gzbar^2(0)}
        &=& \frac{1+\delg -\alpha \,T}{4\,\sqrt{2}\,G_F\,\mmz} \,,
    \label{gzbarfromt}\\[2mm]
      \sbar^2(\mmz)
          &=& \frac{1}{2}
              -\sqrt{\frac{1}{4} -\bar{\alpha}^2(\mmz)
                    \biggl(\frac{4\,\pi}{\gzbar^2(0)} +\frac{S}{4}
                    \biggr)  }\,,
    \label{sbarfroms}\\[2mm]
       \frac{4\,\pi}{\gwbar^2(0)}
             &=& \frac{\sbar^2(\mmz)}{\bar{\alpha}^2(\mmz)}
                -\frac{1}{4}\,(S+U) \,.
    \label{gwbarfromu}
  \end{eqnarray}
  \end{subequations}
It is clear from eqs.~(\ref{gbarfromstu}) that $\gzbar^2(0)$ is 
dependent upon $\delg -\alpha T$, 
$\sbar^2(\mmz)$ is dependent upon $\gzbar^2(0)$,
$\bar{\alpha}(\mmz)$ and $S$, and $\gwbar^2(0)$ is dependent 
upon $\sbar^2(\mmz)$, $\bar{\alpha}(\mmz)$ and $S+U$. 
%
\begin{table}[t]
\caption{\tcaptionofotherdata}
\label{tableofotherdata}\vspace{-2mm}
\tableofotherdata
\end{table}
%
It is instructive to express these form-factors as approximate 
linear combinations of $S$, $T$ and $U$. 
By inserting 
$1/\bar{\alpha}(\mmz)\equiv 4\pi/\ebar(\mmz)=128.72+\delta_\alpha$, 
we find 
  \begin{subequations}
   \label{gbar_approx}
  \begin{eqnarray}
        \gzbar^2(0)    &=& 0.5456 \hphantom{+0.0036\,S}\;\, 
            +0.0040\,\Bigl(T +\frac{0.0055 -\delg}{\alpha}\Bigr) \,,
   \label{gzbar_approx}\\
        \sbar^2(\mmz) &=& 0.2334            +0.0036\,S  
            -0.0024\,\Bigl(T +\frac{0.0055 -\delg}{\alpha}\Bigr)
           \hphantom{ +0.0035\,U }\;\, -0.0026\,\delta_\alpha    \,,
         \qquad
   \label{sbar_approx}\\
        \gwbar^2(0)    &=& 0.4183            -0.0030\,S  
            +0.0044\,\Bigl(T +\frac{0.0055 -\delg}{\alpha}\Bigr)
            +0.0035\,U               +0.0014\,\delta_\alpha      \,.
         \qquad
   \label{gwbar_approx}
  \end{eqnarray}
  \end{subequations}
Here we explicitly retain $\delta_\alpha$ and $\delg$ 
in the expansion. 
The values of the charge form-factors $\gzbar^2(\mmz)$ and $\sbar^2(0)$ 
are then calculated from $\gzbar^2(0)$ and $\sbar^2(\mmz)$ above, 
respectively, by assuming the SM running\cite{hhkm} of the form 
factors between $q^2=0$ and $q^2=\mmz$. 
 
In Table~\ref{tableofotherdata}, we give a list of the data 
from the low-energy neutral-current experiments 
that we use in our analysis. 
They are neutrino-nucleon scattering ($\nu_\mu$--$q$), 
neutrino-electron scattering ($\nu_\mu$--$e$), 
atomic parity violation (APV) and 
polarized electron-deuteron scattering ($e$--D) experiments. 
Additionally, the $W$ mass data\cite{mw93} is given. 
See ref.~\cite{hhkm} for details. 
Also shown are the SM predictions for
$(m_t,\,\mh)=(150,\,100)$, $(150,\,1000)$, 
$(175,\,100)$ and $(175,\,1000)$ in GeV units for 
$\alpha_s=0.116$ and $\delta_\alpha=0$. 

From the low-energy neutral-current experiments 
we obtain two universal parameters, $\gzbar^2(0)$ and $\sbar^2(0)$, 
which may be reparametrized in terms of 
$\gzbar^2(\mmz)$ and $\sbar^2(\mmz)$ by assuming the 
SM running between $q^2=0$ and $q^2=\mmz$. 
In Table~\ref{tableoflencatmz} we show the two universal 
parameters, $\gzbar^2(\mmz)$ and $\sbar^2(\mmz)$, 
determined from $\nu_\mu$--$f$ and $e$--$q$ sectors 
and from all four experiments. 
We used $m_t=175\gev$ and $\mh=100\gev$ when calculating 
the SM running, but the results are insensitive to these values 
in the region $m_t>100\gev$ and $\mh>100\gev$. 
%
\begin{table}[t]
\caption{\tcaptionoflencatmz}
\label{tableoflencatmz}\vspace{-2mm}
\tableoflencatmz
\end{table}
%

\begin{figure}[b]
\begin{center}
\leavevmode\psfig{file=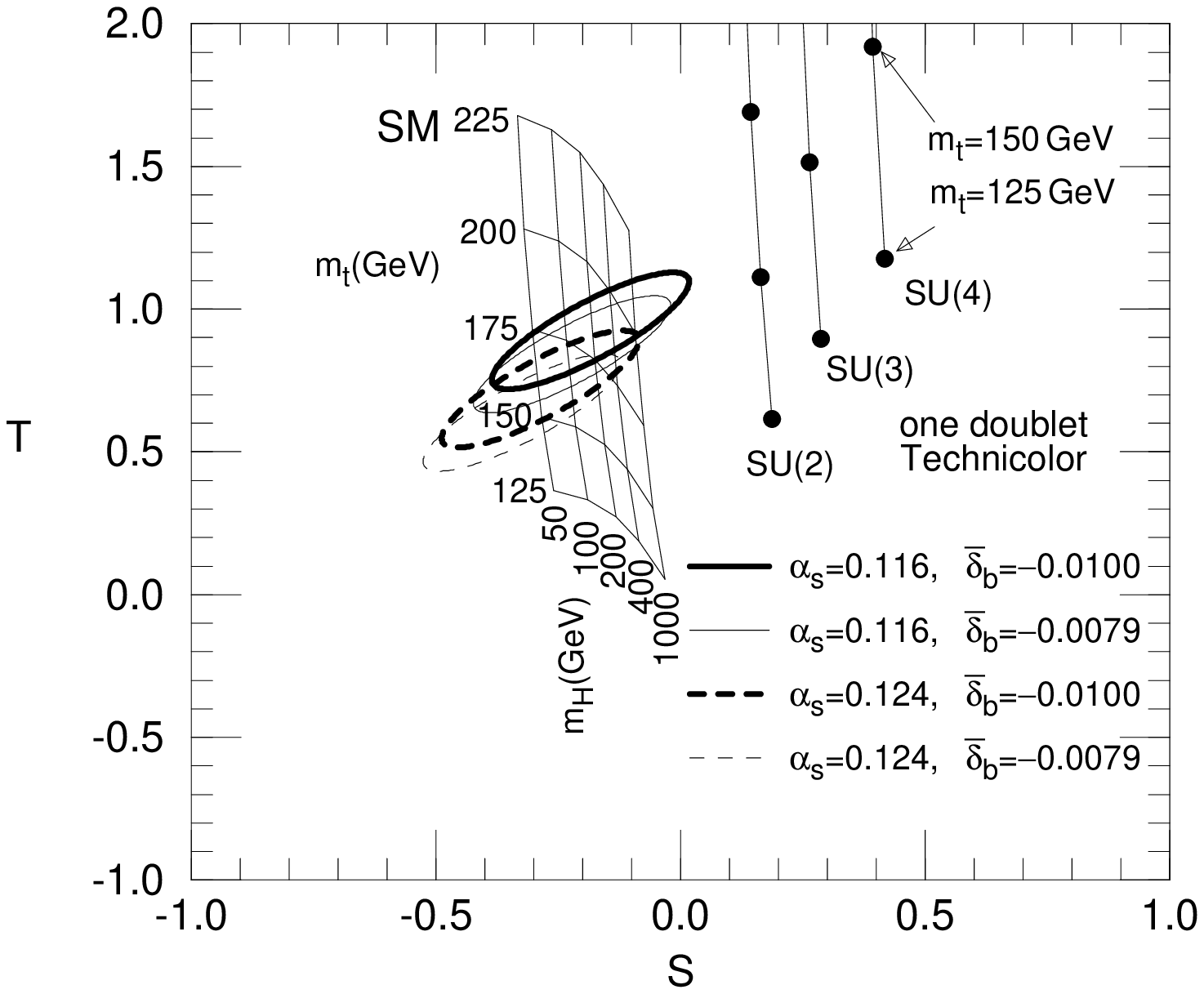,width=12cm,silent=0}
\end{center}\vspace{-7mm}
\caption{\fcaptionofstu}\label{figureofstu}
\end{figure}

The universal electroweak parameter, $\gwbar^2(0)$, 
is obtained from eq.~(\ref{gf}) by combining the data on $\mw$ 
with the muon life-time parameter, $G_F$. We find
%
\fitofmw
%

Now we may combine all the electroweak data; eq.~(\ref{fitofgzbsb}), 
Table~\ref{tableoflencatmz} and eq.~(\ref{fitofmw}). 
Treating $\delb$, $\alpha_s$ and $\delta_\alpha$ 
as external parameters,  and by setting $\delg=0.0055$, 
we find from the three-parameter fit~:
%
\fitofstu
%
The dependence of the $S$ and $U$ parameters upon $\delta_\alpha$ 
may be understood from eq.~(\ref{gbar_approx}). 
For an arbitrary value of $\delg$ the fitted value of $T$ should be 
shifted by $(\delg-0.0055)/\alpha$. 
It should be noted that the uncertainty in $S$ 
coming from $\delta_\alpha=0\pm 0.1$ is 
of the same order as from the uncertainty in $\alpha_s$; 
they are not negligible when compared to the overall error.
The $T$ parameter has little $\delta_\alpha$ dependence, 
but it is sensitive to $\alpha_s$. 

The above results are shown in Fig.~\ref{figureofstu} 
by the 1-$\sigma$ contours in the ($S,\,T$) plane. 
Four cases are shown~: 
$\alpha_s=0.116$ (solid lines) and 0.124 (dashed lines) with 
$\delb=-0.0100$ (thick lines) and $-0.0079$ (thin lines). 
As for the running of the charge form-factors $\sbar^2(q^2)$ 
and $\gzbar^2(q^2)$ between $q^2=0$ and $q^2=\mmz$, 
we set $\mh=100\gev$, and use values of $m_t$ 
corresponding to the above values of $\delb$ 
($m_t=175\gev$ for $\delb=-0.0100$ and 
 $m_t=150\gev$ for $\delb=-0.0079$). 
The SM predictions are also shown in Fig.~\ref{figureofstu} 
by lattices in the region $125\gev<m_t<225\gev$ 
and $50\gev<\mh<1000\gev$. 
The estimates\cite{stu} of $S$ and $T$ for the minimal (one-doublet) 
SU($N_c$) Technicolor (TC) models with $N_c=2,3,4$ 
are also shown in the figure. 
It is clearly seen that the current experiments provide a 
fairly stringent constraint on the simple TC models 
if a QCD-like spectrum and the large $N_c$ scaling 
are assumed\cite{stu}. 
Only with a positive value for $\delta_\alpha$ and a small 
value of $\alpha_s'\equiv \alpha_s +1.6\,\delb$ can 
the $N_c=2$ one-doublet TC model be made consistent with the data. 

\begin{table}[t]
\caption{\tcaptionofstunew}
\label{tableofstunew}\vspace{-2mm}
\tableofstunew
\end{table}

For definiteness we provide, in Table~\ref{tableofstunew}, 
the values of $S$, $T$ and $U$ after the SM contributions are 
subtracted ($S_{\rm new}\equiv S-S_{\rm SM}$, etc.).
The dependence upon $m_t$ and $\mh$ of $S$, $T$ and $U$
appears since we have assumed the SM running 
of the charge form-factors which in turn depends slightly 
on $m_t$ and $\mh$; also because we have used the SM prediction 
for $\delb$ which depends strongly on $m_t$. 
All values in the table are obtained by 
setting $\alpha_s=0.116$ and $\delta_\alpha=0$. 
The values for different values of $\alpha_s$ and $\delta_\alpha$ 
together with the error correlation matrix
can be `read-off' from eq.~(\ref{fitofstu}).


\begin{figure}[b]
\begin{minipage}{16cm}
\begin{center}
\leavevmode\psfig{file=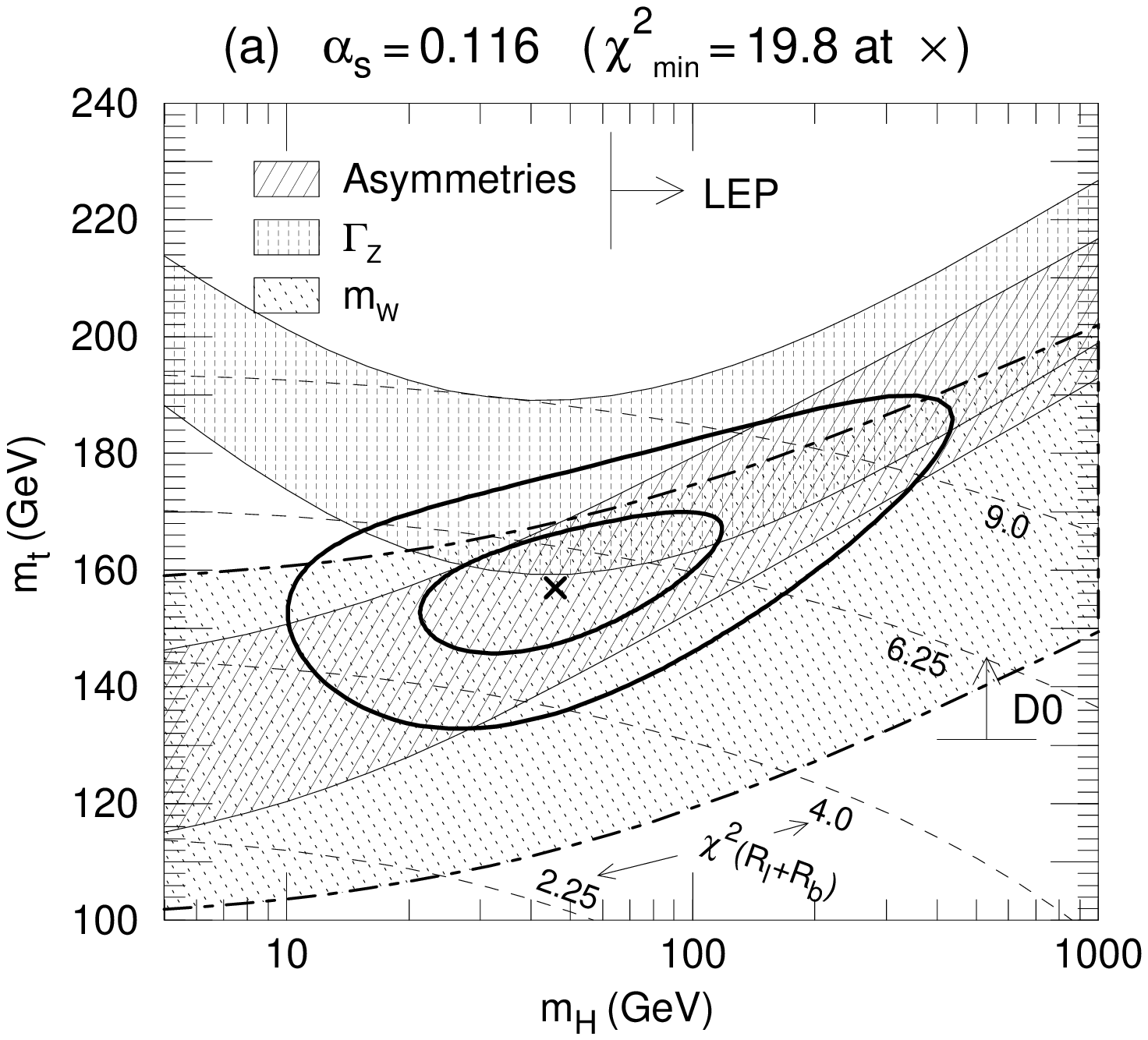,width=7.8cm,silent=0}
\leavevmode\psfig{file=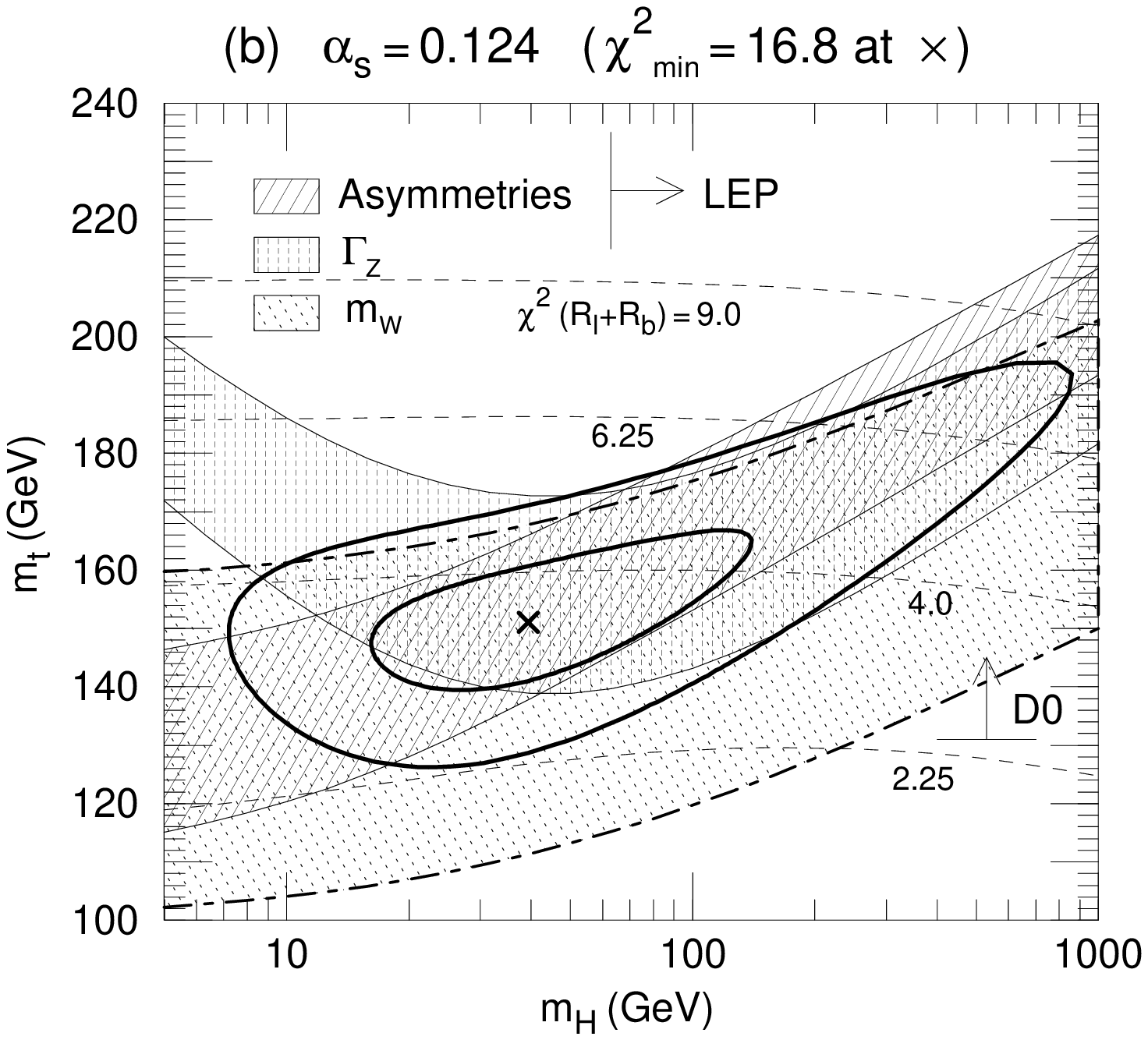,width=7.8cm,silent=0}
\end{center}
\end{minipage}
\caption{\fcaptionofmtmh}\label{figureofmtmh}
\end{figure}

Finally we discuss the constraints on $m_t$ and $\mh$ 
from all the data in Tables~\ref{tableofdata} and \ref{tableofotherdata} 
in the minimal SM. In this case all the form-factors, 
$\gzbar^2(\mmz)$, $\sbar^2(\mmz)$, $\gzbar^2(0)$, $\sbar^2(0)$, 
$\gwbar^2(0)$ and $\delb(\mmz)$, depend uniquely 
on the two mass parameters, $m_t$ and $\mh$. 
Consequently the results of the fits for 
these form-factors constraint $m_t$ and $\mh$. 

Fig.~\ref{figureofmtmh} shows the result of the global 
fit to all electroweak data in the ($\mh,\,m_t$) 
plane 
for $\alpha_s=0.116$ and 0.124 with $\delta_\alpha=0$. 
Here $\alpha_s=0.116$ is the mean value of the PDG 
listing\cite{pdg94} and $\alpha_s=0.124$ 
is the best SM fit value to 
all electroweak data in our three-parameter fit 
in terms of  $m_t$, $\mh$ and $\alpha_s$. 
The thick inner and outer contours correspond to 
$\Delta\chi^2\equiv\chi^2-\chi^2_{\rm min}=1$ ($\sim$ 39\% CL), 
and $\Delta\chi^2=4.61$ ($\sim$ 90\,\%~CL), respectively. 
The minimum of $\chi^2$ is 
indicated by an ``$\times$''. 
The corresponding value of $\chi^2_{\rm min}$ is 
19.8 for $\alpha_s=0.116$ (a) and 16.8 for $\alpha_s=0.124$ (b). 
We also give the separate 1-$\sigma$ constraints arising from 
the $Z$-pole asymmetries, $\Gamma_Z$, and $\mw$. 
The asymmetries constrain $m_t$ and $\mh$ through $\sbar^2(\mmz)$ 
while $\Gamma_Z$ constrains them through the three form-factors 
$\gzbar^2(\mmz)$, $\sbar^2(\mmz)$ and $\delb(\mmz)$. 
In other words, the asymmetries measure the combination of 
$S$ and $T$ as in eq.~(\ref{sbar_approx}); 
both of $S$ and $T$ are functions of $m_t$ and $\mh$. 
On the other hand, $\Gamma_Z$ measures a different combination 
of $S$ and $T$ with an additional constraint from $\delb$. 
A remarkable point apparent from Fig.~\ref{figureofmtmh} 
is that, in the SM, when $m_t$ and $\mh$ are much larger than $\mz$, 
$\Gamma_Z$ depends upon almost the same combination of $m_t$ and $\mh$ 
as the one measured through $\sbar^2(\mmz)$. 
This is because $\gzbar^2(\mmz)$ has a quadratic 
dependence on $m_t$ which is positive while the 
quadratic dependence on $m_t$ of $\delb$ is negative, 
and these two effects largely cancel. 
The result is that the ratio of dependences 
of $\Gamma_Z$ on $m_t$ and $\mh$ are similar to the case of  
of $\sbar^2(\mmz)$. 
Because of this only a band of $m_t$ and $\mh$ can be 
strongly constrained from the asymmetries and $\Gamma_Z$ alone, 
despite their very small experimental errors. 
The constraint from the data on $\mw$ overlaps this allowed region. 

\begin{figure}[b]
\begin{minipage}[t]{7.8cm}
\leavevmode\psfig{file=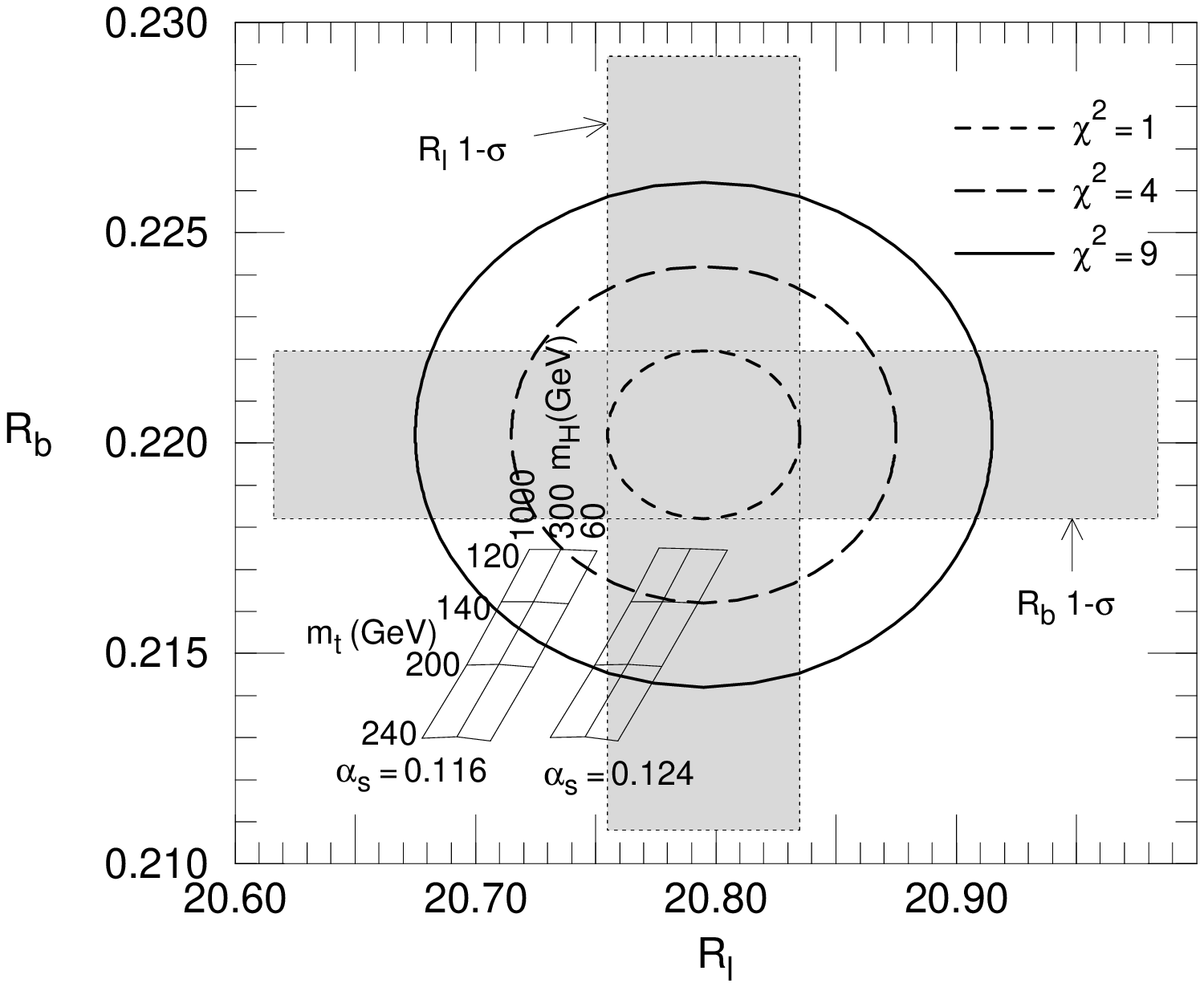,width=7.8cm,silent=0}\vspace{-7mm}
\flushright{
  \begin{minipage}[t]{7cm}
     \caption{\fcaptionofrlrb}\label{figureofrlrb}
  \end{minipage}
}
\end{minipage}
\hfill
\begin{minipage}[t]{7.8cm}
\leavevmode\psfig{file=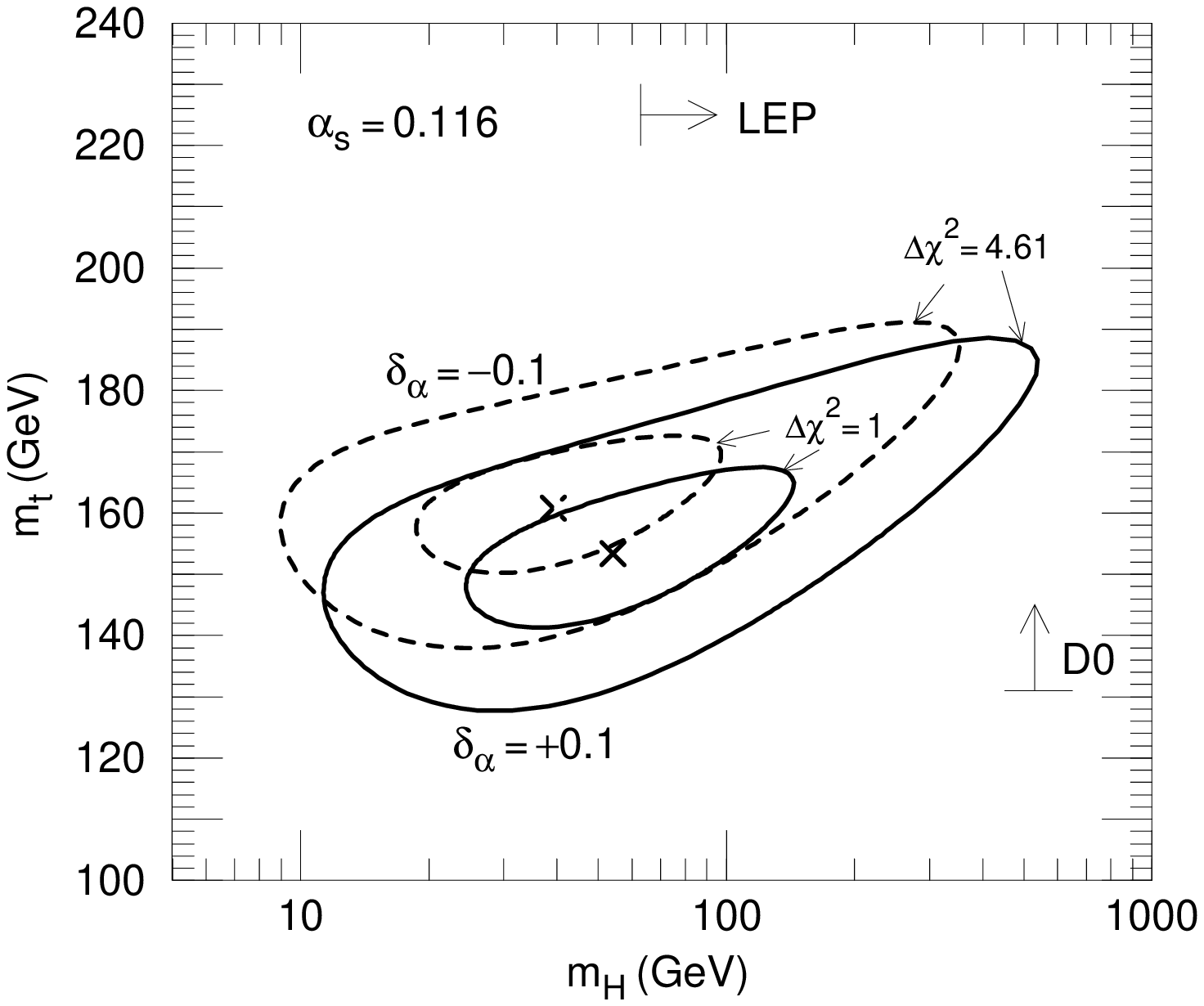,width=7.8cm,silent=0}\vspace{-7mm}
\flushright{
  \begin{minipage}[t]{7cm}
     \caption{\fcaptionofmtmhda}\label{figureofmtmhda}
  \end{minipage}
}
\end{minipage}
\end{figure}

Quantities which help to disentangle 
the above $m_t$-$\mh$ correlation are $R_\ell$ and $R_b$. 
The constraints from these two data are shown in 
Fig.~\ref{figureofmtmh} by dashed lines corresponding to 
$\sqrt{\chi^2}=1.5,\,2.0\,,2.5\,,3.0$. 
As shown in Fig.~\ref{figureofrlrb} the SM prediction for $R_\ell$ 
is very sensitive to the assumed value of $\alpha_s$, and, 
for $\alpha_s=0.116$, the data favors the region where both $m_t$ 
and $\mh$ are small; for $\alpha_s=0.124$ 
it does not provide a stringent limit on $m_t$ and $\mh$. 
This explains the difference between 
the two cases, $\alpha_s=0.116$ (Fig.~\ref{figureofmtmh}a) 
and $0.124$ (Fig.~\ref{figureofmtmh}b). 
On the other hand, data on $R_b$ favors small $m_t$ 
almost independently of $\mh$. 
It is hence the $R_\ell$ and $R_b$ data that constrain 
the values of $m_t$ and $\mh$ from above. 
If not for the data on $R_\ell$ and $R_b$, 
the common shaded region in Fig.~\ref{figureofmtmh} with 
very large $\mh$ $(\mh\sim 1\tev)$ could not be excluded by the 
electroweak data alone. 

The results of the fits for different values of $\delta_\alpha$  
are shown in Fig~\ref{figureofmtmhda}. 
The case for $\delta_\alpha=-0.1$ is shown by dashed lines, 
and the case for $\delta_\alpha=+0.1$ is shown by solid lines. 
In both cases the inner and outer contours correspond to 
$\Delta\chi^2=1$ ($\sim$ 39\% CL), 
and $\Delta\chi^2=4.61$ ($\sim$ 90\,\%~CL), respectively.

If the lower bound for $\mh$ ($\mh>63\gev$ at 95\%~CL) measured 
by the LEP experiments\cite{mh_limit} 
is imposed then $m_t$ below $130\gev$ is clearly 
disfavored for these regions on $\alpha_s$ and $\delta_\alpha$.  
This agrees with the directly established lower top-mass 
limit\cite{top_cdf_limit,top_d0}.

The $\chi^2$ function of the global fit to all electroweak 
data can be parametrized in terms of the four parameters 
$m_t$, $\mh$, $\alpha_s$ and $\delta_\alpha$ 
together with the constraint 
$\delta_\alpha=0.0 \pm 0.1$\cite{piqq_h_latest} by~:
%
\chisqsm
%
with
%
\fitofmt
%
and 
%
\chisqhsm
%
Here $m_t$ and $\mh$ are measured in GeV. 
This parametrization reproduces the exact $\chi^2$ function 
within a few percent accuracy in the range $100\gev<m_t<250\gev$, 
$60\gev<\mh<1000\gev$ and $0.10<\alpha_s(\mz)<0.13$. 
The best-fit value of $m_t$ for a given set of $\mh$, 
$\alpha_s$ and $\delta_\alpha$ is readily obtained from 
eq.~(\ref{fitofmtbest}) with its approximate error of
(\ref{fitofmterror}). 
For $\mh=60,300,1000\gev$, $\alpha_s=0.116$ and $\delta_\alpha=0$, 
one obtains 
  \begin{eqnarray}
     m_t = 
     \left\{
     \begin{array}{llll}
       159 \pm 9  \gev & \mbox{for} & \mh=60   \gev 
       &(19.9/19)\\
       180 \pm 9  \gev & \mbox{for} & \mh=300  \gev 
       &(23.1/19)\\
       199 \pm 8  \gev & \mbox{for} & \mh=1000 \gev
       &(27.8/19)\\
     \end{array}
     \right.  \,,
  \end{eqnarray}
where $(\chi^2_{\rm min}/{\rm d.o.f.})$ is shown in brackets. 
We note here that $\chi^2_{\rm min}=27.8$ for $\mh=1000\gev$ 
reflects the discrepancies between the fitted value of $m_t$ and 
the $R_\ell$ and $R_b$ data. 
One can observe from eq.~(\ref{fitofmt}) that changing $\alpha_s$ by
$\pm 0.005$ shifts the best-fit values of $m_t$ about $\mp 3\gev$, 
while changing $\delta_\alpha$ by $\pm 0.1$ shifts it about $\mp 5\gev$. 

Due to its quadratic form it is easy to obtain from 
eq.~(\ref{total_chisqsm}) results which are independent of 
$\alpha_s$ and/or $\delta_\alpha$. 
Also, additional constraints on the external parameters 
$\alpha_s$ and $\delta_\alpha$, such as those from their improved 
measurements or the constraint from 
the grand unification of these couplings 
may be discussed without difficulty. 


In view of the recent publication by the CDF 
collaboration\cite{top_cdf} concerning evidence for 
the top quark with $m_t=174\pm 16\gev$, 
it is instructive to anticipate the impact a precise 
measurement of the top-quark mass would have in the context of 
the present electroweak data. 
%
\begin{figure}[b]
\begin{center}
\leavevmode\psfig{file=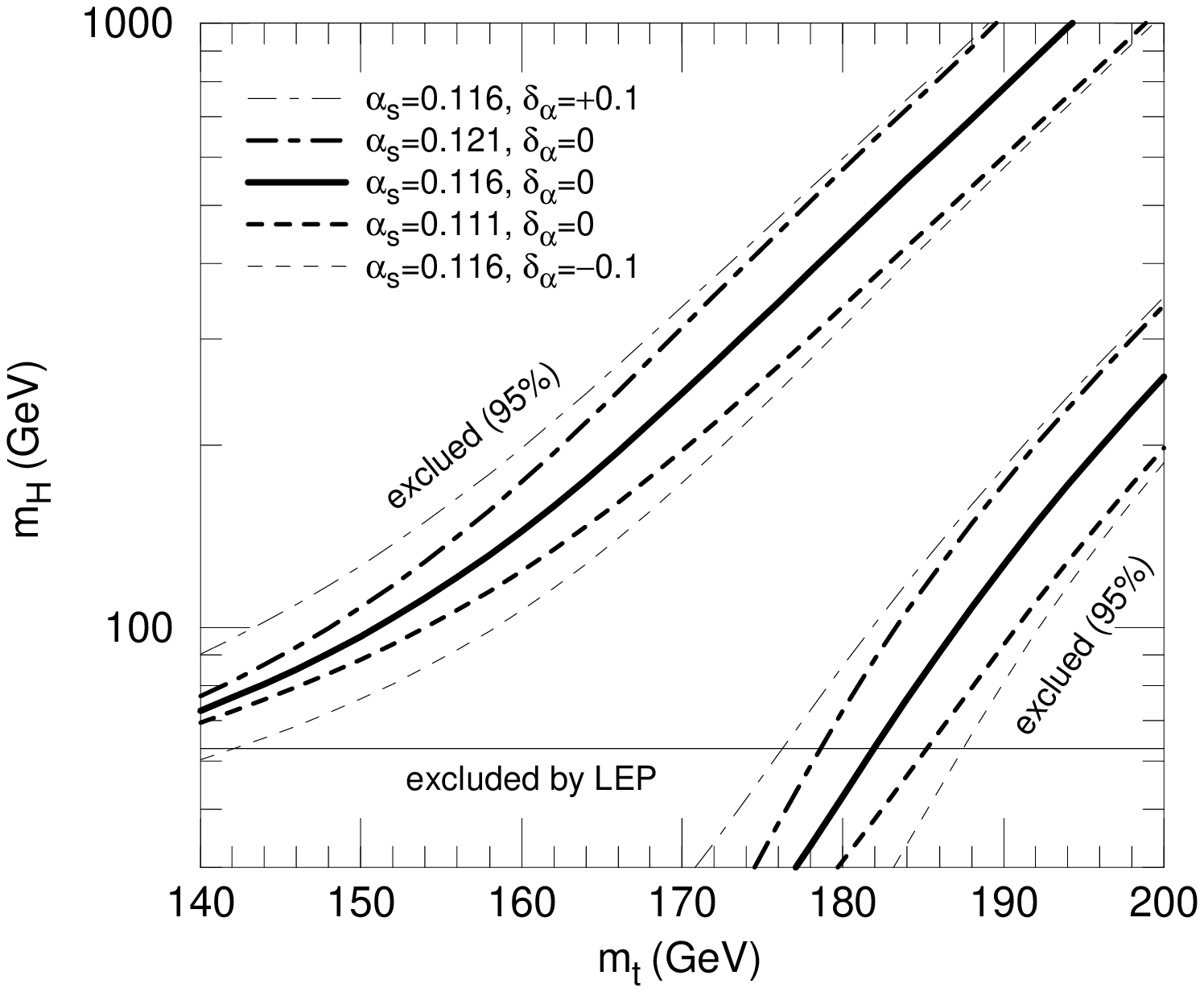,width=10cm,silent=0}
\end{center}\vspace{-7mm}
\caption{\fcaptionofmhcl}\label{figureofmhcl}
\end{figure}
In the discussion below 
we treat $m_t$ as an external parameter, 
and hence we discuss the sensitivity of the present electroweak 
data to $\mh$ while assuming that $m_t$ is known precisely. 
The 95\%~CL upper/lower bounds on $\mh$ from the electroweak data 
are shown in Fig.~\ref{figureofmhcl} as functions of $m_t$. 
Shown by the thick solid lines are the bounds for 
$\alpha_s(\mz)=0.116$ and $\delta_\alpha=0$. 
The bounds for $\alpha_s(\mz)=0.111$ and 
$0.121$ with $\delta_\alpha=0$ 
are shown by dashed and dot-dashed thick lines, respectively, 
while the bounds for $\delta_\alpha=-0.1$ and $+0.1$ 
with $\alpha_s=0.116$ 
are shown by dashed and dot-dashed thin lines, respectively. 
In the region $160\gev <m_t< 190\gev$ the upper bound on $\mh$
at the 95\%~CL is approximately expressed as%
\footnote{\normalsize \baselineskip 18pt
  One comment is in order. 
  Although our approximate formulae for the $\chi^2$ of the SM fit, 
  (\ref{total_chisqsm}), reproduce the   exact result within about 
  1\% accuracy in the Higgs-mass range   $63\gev<\mh <1000\gev$, 
  one should not use these formulae to find the confidence levels 
  of $\mh$ for small $m_t$; the neighborhood of the minimum 
  of the $\chi^2$ is outside the above range,  and in this case 
  the exact $\chi^2$ and the approximate formulae are 
  significantly different. See ref.~\cite{hhkm} for discussions. 
}
  \begin{eqnarray}
    \ln\frac{\mh}{100} < 
     \left\{\begin{array}{c}
        0.91 \\1.13 \\ 1.38
     \end{array} \right\}
     + 
     \left\{\begin{array}{c}
        0.85\\ 0.91\\ 0.95
     \end{array} \right\}\,
     \frac{m_t-174}{16} +0.34\,\frac{\delta_\alpha}{0.1}
     \quad\mbox{for}\quad \alpha_s = 
      \left\{\begin{array}{l}
        0.111\\ 0.116\\0.121
      \end{array} \right. \,,
  \end{eqnarray}
where $m_t$ and $\mh$ are measured in GeV. 
For a smaller value of $m_t$, 
a rather stringent upper bound on $\mh$ is obtained. 
Since these bounds are very 
sensitive to the value of $m_t$ as well as the assumed 
values of $\alpha_s$ and $\delta_\alpha$ 
a further, more accurate measurement of $m_t$ will give 
more definite information on $\mh$. 


To summarize: 
We have performed a comprehensive analysis 
of the recent electroweak data at LEP/SLC. 
The two universal parameters, $\gzbar^2(\mmz)$, $\sbar^2(\mmz)$, 
and the $\zbb$ vertex form-factor, $\delb(\mmz)$, are 
determined from these data.
The $S$, $T$ and $U$ parameters are also determined 
by including the data from low-energy neutral-current experiments 
and $W$-mass data. 
The errors in $S$ and $T$ are much reduced from 
those of the previous analysis\cite{hhkm}, 
and simple TC models are clearly disfavored. 
As for the SM fit, the value of $m_t$ favored by the electroweak 
data is in good agreement with the value favored by CDF\cite{top_cdf}. 
We also note that an upper bound on the Higgs-boson mass 
can be obtained for a given value of $m_t$, 
and that a stringent upper bound ($\mh<140\gev$) is found 
for rather small $m_t(\simlt 160\gev)$. 
At all stages of our analysis, we have discussed, in detail, 
the uncertainties coming from 
the QCD coupling strength, $\alpha_s$, and 
the shift $\delta_\alpha \equiv 1/\bar{\alpha}(\mmz)-128.72$ 
by presenting all fit results as functions 
of $\alpha_s$ and $\delta_\alpha$. 
The improvement of the measurements of these parameters 
are crucial to the search for physics beyond 
the SM through radiative corrections. 

\vspace{3mm}
\subsection*{Acknowledgements}

The author wishes to thank B.K.~Bullock, K.~Hagiwara, D.~Haidt, 
R.~Szalapski and Y.~Yamada for clarifying discussions. 
He is also grateful to the KEK theory group for extending 
its hospitality during the course of this work. 


\end{document}